%% file: main.tex
\documentclass[10pt]{extarticle}
\usepackage[utf8]{inputenc}
\usepackage[normalem]{ulem}% To strikethrough some text
\usepackage[]{amsmath} %felqn
\usepackage{amssymb}
\usepackage{nccmath} % temporarilly centered equation
\usepackage{gensymb} % needed for \degree
\usepackage{bm}
\usepackage{colortbl}  % cellcolor
\usepackage[table,dvipsnames]{xcolor}    % loads also »colortbl«
\usepackage{graphicx, wrapfig, subcaption, setspace, booktabs}
\usepackage{subcaption} % For subfigures
\usepackage{float}
\usepackage{parskip}
\usepackage{empheq}
\usepackage[theorems,skins]{tcolorbox}
\usepackage{xcolor}
\usepackage{tikz}
\usepackage{natbib}
\usepackage{wrapfig}   % Für die wrapfigure-Umgebung
\usepackage{comment}
\usepackage[a4paper, total={14cm, 24cm},]{geometry}
\usepackage{layout}    %  um die Seitenränder als Bild auszugeben
\usepackage{empheq}
\usepackage{xspace} % xspace used in glossaries
\usepackage{myglossaries}
\usepackage[colorlinks = true,
            linkcolor = blue,
            urlcolor  = blue,
            citecolor = blue,
            anchorcolor = blue]{hyperref}
\begin{document}

% =============================================================================
% =============================================================================
%   TITLE
% =============================================================================
% =============================================================================
\title{\Large\textbf{Density-ratio invariance of the optimal passive pitching 
\\ of a self-propelled three-dimensional flapper}}
\author{Maximilian Schenk, Manuel Garc\'ia-Villalba and Manuel Moriche}
\date{\vspace{-2ex}}
\maketitle
\normalsize
\begin{center}
\textit{Institute  of Fluid Mechanics and Heat Transfer, 
TU Wien, Vienna, Austria} \\
\end{center}

% =============================================================================
% =============================================================================
%   ABSTRACT
% =============================================================================
% =============================================================================
\rule{\textwidth}{0.6pt}
\vspace{-0.2cm}

\begin{abstract}
We study numerically a three-dimensional flapper of finite span that propels
itself by a prescribed sinusoidal heaving motion of its centre of mass.
The pitching motion of the rigid flapper results from the balance between a 
torsional spring, the flapper's inertia and the fluid forces.
We vary the stiffness of the spring for three solid-to-fluid density ratios, 
$\gls{rhor}=1$, $10$ and $100$.
For all three density ratios, the propulsive speed reaches approximately
the same maximum, while the stiffness at which it occurs increases with 
the density ratio.
When the stiffness is shifted by the respective optimum, the propulsive speeds 
of the three density ratios collapse onto a single curve.
At the optimum, the three flappers pitch with the same amplitude and lead the
heaving motion by $113^\circ$, so that they experience the same effective 
angle of attack and the same vertical hydrodynamic force.
The passive pitching acts as a feathering motion that turns the flapper
towards the incoming flow.
Stiffer springs feather too little, so the effective angle of attack remains
large, the flow separates at the leading and trailing edges, and the
hydrodynamic force acts mainly in the heaving direction.
Softer springs result in excessive and delayed pitching of the flapper, and the
benefits of feathering are outweighed by weaker hydrodynamic forces.
The optimum balances these two limits.
The input power increases strongly with the density ratio because of the inertia 
of the flapper, so that the distance travelled per unit input work decreases, 
whereas the propulsive efficiency, which accounts for the kinetic energy of
the flapper, shows the opposite trend.
A comparison with the resonant stiffness of the flapper considering the added
inertia of quiescent fluid, shows that the optimal spring is stiffer than the
resonant one for all density ratios, by a nearly constant amount.
The optimal spring is therefore not set by resonance but by the feathering 
motion it produces, and a flapper of higher density ratio only needs a stiffer 
spring to reach it.
\end{abstract}
\rule{\textwidth}{0.6pt}

\input{tex/01_introduction}
\input{tex/02_methodology.tex}

\input{tex/03_results.tex}
\input{tex/04_conclusion}

% =============================================================================
% =============================================================================
%   OUTRO
% =============================================================================
% =============================================================================
\section*{Statements and Declarations}

\textbf{Authors’ Individual Contributions}
Conceptualization: MGV, MM;
Methodology: MGV, MM;
Investigation: MS;
Formal analysis: MS;
Visualization: MS;
Software: MS, MM;
Writing – original draft preparation: MS;
Writing – review and editing: MGV, MM;
Supervision: MM.

\textbf{Statement on the use of generative AI}
 During the preparation of this work, the authors used Claude (Anthropic) in 
 order to improve the readability and flow of the text as well as to check for 
 consistency of the text. 
 After using this tool, the authors reviewed and edited the content as needed 
 and take full responsibility for the content of the published article.

%\textbf{Acknowledgments} 

\clearpage
\bibliographystyle{apalike}   
\bibliography{reference.bib}

\clearpage
\appendix
\input{tex/05_appendix}

\end{document}

%% file: tex/01_introduction.tex
% =============================================================================
% =============================================================================
%   INTRODUCTION
% =============================================================================
% =============================================================================
\section{Introduction}
\label{sec:introduction}

% ------------------------------------------------------------------------------
% MOTIVATION - PASSIVE PITCHING
Flapping-based propulsion is central to the locomotion of many aquatic and 
aerial organisms, and it has motivated a sustained engineering effort to
build propulsors for underwater and aerial vehicles
\citep{triantafyllou2000,shyy2010,wang2016,smits2019}.
Early work established that distributed chordwise flexibility can improve
propulsive efficiency \citep{katz1978}, and a linear two-dimensional theory 
later showed that concentrating this flexibility in a torsional spring at the 
leading edge maximises the thrust \citep{moore2015}.
Many flapping propulsors actuate only the heaving motion, and let the pitching 
motion emerge passively from the fluid--structure interaction, typically 
through an elastic hinge or a torsional spring 
\citep{spagnolie2010,zhang2010,chen2016,hu2021}.
The pitching motion then results from the balance between the hydrodynamic
torque, the inertia of the flapper and the spring response.
Its amplitude and its phase relative to the heaving motion therefore depend on 
both the spring stiffness and the density of the flapper relative to the fluid.
For prescribed kinematics, this phase alone strongly affects the thrust and 
efficiency of a heaving and pitching airfoil \citep{Guglielmini2004,moriche2017}.

% ------------------------------------------------------------------------------
% CANONICAL SELF-PROPELLED PROBLEM
In self-propelled locomotion, the forward speed results from the hydrodynamic
force generated by flapping motion.
In the canonical model of this problem, a plate or foil heaves with a prescribed 
motion and is free to move in a direction perpendicular to the imposed heaving 
(streamwise direction in the case of successful self-propulsion).
Even in the absence of pitching motion, a rigid foil can propel itself when its
flapping frequency is above a critical threshold, thanks to a symmetry-breaking
instability of the flow \citep{vandenberghe2004,alben2005,deng2018}.
If a torsional spring is added at the leading edge of the foil, the passively
pitching foil can swim considerably faster \citep{spagnolie2010,zhang2010}.
\citet{zhang2010} showed that the direction of locomotion depends on the ratio 
of the natural frequency of the spring-plate system, which is set by the spring 
stiffness, relative to the heaving frequency.
The plate moves forward when the heaving frequency is below its natural frequency, 
showing an optimum when the flapping frequency is approximately half of the 
natural frequency.
\citet{arora2018} extended this model to a chordwise-flexible plate, represented 
by several rigid segments connected by torsional springs.
The highest cruising velocity again occurs at a heaving frequency below the
natural frequency.
This optimum lies further above it as the plate inertia decreases relative to the 
hydrodynamic forces.
\citet{arranz2022,arranz2022_multibody} extended this model to a three-dimensional
plate of finite span.

% ------------------------------------------------------------------------------
% RESONANCE
In these studies, the natural frequency lies above the optimal heaving frequency, 
which implies that the optimal spring is stiffer than the resonant one.
However, it is reasonable to expect that the best performance is given by the
combination of stiffness and inertia that results in resonance of the pitching
motion.
Indeed, \citet{spagnolie2010} found the highest propulsive speed close to 
resonance, and \citet{phoemsapthawee2020} related the highest efficiency of
heaving foils with a torsional spring to resonance.
In a linear potential-flow model of a self-propelled foil elastically mounted
through translational and torsional springs, \citet{lopez2023} likewise 
associated the best swimming performance with resonant spring constants.
\citet{xiao2014} found the largest thrust for their softest spring, close to
resonance, but the highest efficiency for a stiffer spring, with a natural
frequency two to three times the flapping frequency.
In contrast, experiments with flexible wings show the best performance for wings 
stiffer than resonant \citep{thiria2010,ramananarivo2011}, and the linear theory 
of \citet{moore2015} also gives the largest thrust for a spring stiffer than the 
resonant one.
In a linear slender-body model of a flexible plate with active heave and 
passive pitch, \citet{Wu2020} found that resonance mainly enhances the thrust, 
whereas the phase dynamics dominates the propulsive efficiency.
In the linear inviscid limit, \citet{fernandez2021} found the best performance
of a heaving foil with passive pitching usually close to the first natural
frequency of the coupled fluid--structure system.
These results show that whether the optimal stiffness coincides with resonance
remains an open question.

% ------------------------------------------------------------------------------
% DENSITY
The resonant stiffness depends on the inertia of the flapper, which makes the 
density ratio between the flapper and the fluid, \gls{rhor}, a key parameter of 
the problem.
It sets the inertia of the flapper relative to the added inertia of the fluid 
and to the hydrodynamic forces, and it therefore affects both the amplitude and 
the phase of the passive pitching motion.
At fixed spring stiffness, \citet{shao2016} found that a foil of lower density 
ratio reverses its direction of locomotion at a higher flapping frequency.
This shift is consistent with a reversal near the natural frequency of the foil 
\citep{zhang2010}, but the agreement deteriorates as the density ratio decreases.
For a wing of infinite span, \citet{xiao2014} found that a larger inertia leads to 
a slightly higher propulsive speed with a smaller pitching amplitude.
\citet{arora2018} compared two-dimensional plates at the optimal stiffness of
each case, varying the inertia of the plate through the product of the density 
ratio and the plate thickness.
They found that, as the inertia of the plate increases, its maximum propulsive 
speed decreases and its input power generally increases.
These studies keep the spring stiffness, the natural frequency or the optimal 
stiffness fixed as the density ratio varies, which makes their trends difficult 
to compare.

% ------------------------------------------------------------------------------
% 3D
One common assumption in the above-mentioned studies is two-dimensional flow,
whereas real flappers are influenced i) by the potential development of
three-dimensional flow structures and ii) by geometrical effects from having
a finite span.
At low aspect ratios, the tip vortices connect the shed vortices from the
leading and trailing edge into rings, and the wake splits into two branches
\citep{dong2006,kurt2020}.
This three-dimensional evolution of the wake changes the forces on the flapper, 
and two-dimensional simulations cannot capture it \citep{arranz2020,zurman2020}.
Three-dimensional wings with torsional springs have been studied while held
in a free stream \citep{hu2021,martinezmuriel2023,martinezmuriel2024,hu2025}.
Self-propelled three-dimensional flappers with passive pitching have been
modelled as continuously flexible plates \citep{tang2016,yeh2016,wang2020} or 
as plates of rigid segments connected by torsional springs 
\citep{arranz2022,arranz2022_multibody}.
In the latter studies, the density ratio is fixed, and the stiffness is either 
taken from the two-dimensional optimum or chosen from two values.
\citet{xiao2014} varied the density ratio of a self-propelled wing with a
torsional spring, and found that its effect depends on the aspect ratio.
Their wing, however, pitched about its mid-chord, and the density ratios were
compared at a fixed frequency ratio rather than at the optimal stiffness.
Outside self-propelled locomotion, two studies relate the inertia of the 
flapper to its optimal stiffness.
\citet{phoemsapthawee2020} compared three-dimensional foils of different
inertia at the stiffness of maximum efficiency, but with a prescribed advance
velocity and a boundary element method, which neglects viscosity.
In hovering flight, \citet{dai2012} found that the optimal stiffness for lift
of an elastic wing depends on its mass ratio.
To our knowledge, no three-dimensional study has compared self-propelled
flappers of different density ratios at their respective optimal stiffness,
or related these optima to the resonant stiffness.

% -----------------------------------------------------------------------------
% SET-UP AND OBJECTIVES.
The present work addresses this gap by investigating, numerically, how
the density ratio affects the fluid--structure interaction of a 
three-dimensional self-propelled flapper.
The flapper is a rigid body that heaves with a prescribed motion and pitches
passively about its centre of mass, restrained by a torsional spring.
The kinematics follow the configuration of \citet{arranz2022}, and the results 
serve as the reference for a tandem configuration reported elsewhere 
\citep{schenk2026}.
For density ratios from  $\gls{rhor}=1$ to $\gls{rhor}=100$, we determine the 
spring stiffness that maximises the propulsive speed.
We then examine how the density ratio changes the amplitude and phase of
the passive pitching motion at this optimal stiffness.
Finally, we compare the input power and the distance travelled per unit
input work of flappers with different density ratios.

%-------------------------------------------------------------------------------
% ORGANIZATION 
The paper is organised as follows.
Section~\ref{sec:meth} describes the problem, the numerical method and the 
performance indicators.
Section~\ref{sec:results} presents the results.
We first analyse the effect of the spring stiffness for $\gls{rhor}=1$
(\S~\ref{sec:results:kinematics}) and then extend the analysis to the three
density ratios considered (\S~\ref{sec:results:optimum}).
Section~\ref{sec:conclusions} summarises the main findings.

%% file: tex/02_methodology.tex
% =============================================================================
% =============================================================================
%   METHODOLOGY
% =============================================================================
% =============================================================================
\section{Methodology} \label{sec:meth}

% -----------------------------------------------------------------------------
\subsection{Problem description}
\label{sec:meth:prob}
We investigate the self-propulsion of a three-dimensional flapper that pitches 
passively by the response of a torsional spring to hydrodynamic forces.
% =============================================================================
%   FLOW
% =============================================================================
The flapper is submerged in an otherwise quiescent fluid of density \gls{rhof}
and kinematic viscosity \gls{nu}.
We solve the Navier--Stokes equations for an incompressible flow
\begin{subequations} \label{eq:ns}
\begin{align}
   \label{eq:ns/mass}
   \nabla  \cdot \gls{vu} &= 0 \, ,
   \\ \label{eq:ns/mom}
   \ppt{\gls{vu}}
   + \left(\gls{vu} \cdot \nabla \right) \gls{vu} 
       &= -\frac{\nabla \gls{p}}{\gls{rhof}} + \gls{nu} \nabla^2 \gls{vu} \,,
\end{align}
\end{subequations}
where $\gls{vu}=(\gls{vu_x},\gls{vu_y},\gls{vu_z})$ is the fluid velocity and
$\gls{p}$ is the pressure.
The equations in \eqref{eq:ns} are complemented with a no-penetration, no-slip
boundary condition at the surface of the flapper.
% -----------------------------------------------------------------------------

% =============================================================================
%   PARTICLE
% =============================================================================
\begin{figure}[t!]
    \centering
    \includegraphics[width=1.0\textwidth]{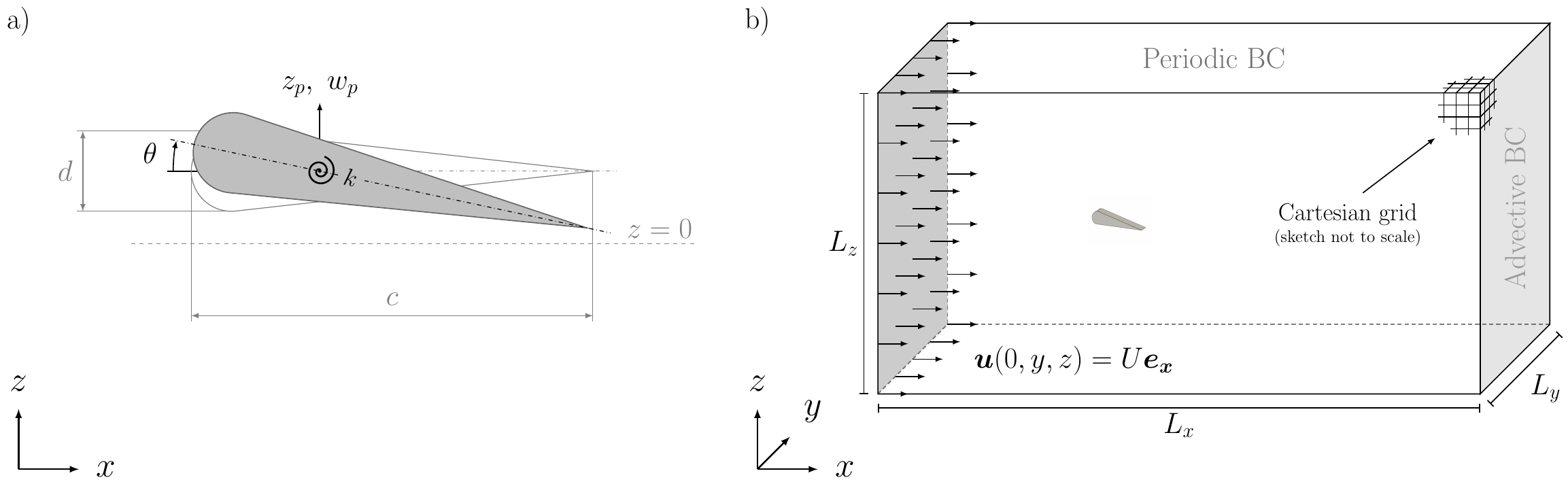}
    \caption{%
    a) Sketch of the problem including the main geometric definitions and
    the imposed motion of the flapper and b) computational domain with the
    boundary conditions (BC).
    The dashed horizontal line in a) marks the domain centreline in the
    vertical direction ($\gls{vx_z}=0$).
    The white flapper in a) represents its equilibrium position relative to
    the spring ($\gls{th} = 0$).}
    \label{fig:meth:sketch}
\end{figure}
%

% -----------------------------------------------------------------------------
We model the flapper as a rigid body of uniform density \gls{rhop}.
Its cross-section is constant along the span \gls{b} and consists of a rounded 
leading edge of diameter \gls{d} and two straight edges tangent to the rounded
leading edge and tapering towards a sharp trailing edge.
The diameter \gls{d} therefore sets the maximum thickness of the flapper.
We choose $\gls{d}=0.2\gls{c}$, which gives a cross-sectional area
$A_c=0.106\,\gls{c;sq}$.
With the span $\gls{b}=0.5\gls{c}$, the volume of the flapper is
$\gls{Vp}=A_c\,\gls{b}=0.053\,\gls{c;cb}$.
The flapper pitches about its centre of mass, \gls{vxp}, located at a
distance of $0.349\,\gls{c}$ from the leading edge.
The specific moment of inertia of the flapper (moment of inertia per unit mass)
around an axis parallel to the span and passing through the centre of mass is
$\gls{J}=0.055\,\gls{c;sq}$.
Appendix~\ref{sec:app:flapper_geometry} gives the derivation of these 
quantities.
% -----------------------------------------------------------------------------

% -----------------------------------------------------------------------------
The position, 
$\gls{vxp} = \left(\gls{vxp_x},\gls{vxp_y},\gls{vxp_z}\right)$,
and velocity,
$\gls{vup} = \left(\gls{vup_x},\gls{vup_y},\gls{vup_z}\right)$,
of the centre of mass are partially constrained.
The lateral direction is inhibited, thus $\gls{vup_y}=0$, and in the vertical
direction we impose a sinusoidal heaving motion
\begin{subequations}\label{eq:meth:imposed}
\begin{align}
    \label{eq:meth:pos} 
    \gls{vxp_z}(t) &= \gls{A} \cos(2\pi \gls{f} t ) \,,
    \\ \label{eq:meth:vel} 
    \gls{vup_z}(t) &= - 2 \pi \gls{f} \gls{A} \sin(2\pi \gls{f} t ) \,,
\end{align}
\end{subequations}
where $\gls{A}$ is the heaving amplitude and $\gls{f}$ the oscillation frequency.
Furthermore, $x/z-$rotations are suppressed, hence the only angular motion
corresponds to pitching (rotation around the span, $y$-direction),
characterized by the pitching angle, \gls{th}, and the angular velocity,
\gls{om} (see sketch in Fig.~\ref{fig:meth:sketch}).
Considering the above, the non-constrained degrees of freedom governed by the 
Newton--Euler equations of motion are reduced to: i) the linear momentum
equation in the streamwise direction
\begin{equation}\label{eq:lin}
    \gls{rhop}\gls{Vp} \, \ddt{\gls{vup_x}} =  \gls{vF_x}\,,
\end{equation}
and ii) the angular momentum equation in the spanwise direction
\begin{equation}\label{eq:mom}
    \gls{rhop}\gls{Vp} \gls{J}\, \ddt{\gls{om}} = \gls{M} + \gls{K} \,,
\end{equation}
where \gls{vF_x} is the \gls{vx_x}-component of the hydrodynamic force
$\gls{vF}=\int_{S} \tau \cdot \vec{n} \, \mathrm{d}\sigma$,
where $\vec{n}$ is the outward unit normal and
$\tau = -\gls{p}\,\bm{I}
+ \gls{rhof}\gls{nu}(\nabla\gls{vu} + (\nabla\gls{vu})^{\top})$
the hydrodynamic stress tensor, with $\bm{I}$ the identity tensor.
\gls{M} is defined as the spanwise component of the hydrodynamic torque
$\gls{vM} = \int_{S} \vec{r} \times (\tau \cdot \vec{n}) \
\mathrm{d}\sigma$, with $\vec{r} = \gls{vx} - \gls{vxp}$ the
position of a surface point relative to the centre of mass.
\gls{K} is the restoring torque of a torsional spring located at \gls{vxp}
(Fig.~\ref{fig:meth:sketch}), which allows a passive pitching motion
mimicking chordwise flexibility \citep[as in][]{arranz2022}.
Assuming a constant stiffness \gls{k}, this torque depends linearly on the
pitching angle \gls{th},
\begin{equation}
    \gls{K} = - \gls{k} \gls{th} \,.
    \label{eq:meth:K}
\end{equation}
Gravity is not considered since the vertical motion of the flapper is imposed.
% -----------------------------------------------------------------------------

% =============================================================================
%   PARAMETER
% =============================================================================
%
With the geometry of the flapper fixed, the problem is governed by four
non-dimensional numbers.
Here, we use the density ratio between the solid and the  fluid,
\begin{equation}
\gls{rhor}=\frac{\gls{rhop}}{\gls{rhof}} \,,
\end{equation}
the Reynolds number based on the maximum vertical speed of the centre of mass,
$\gls{V} = 2\pi\,\gls{A}\,\gls{f}$,
\begin{equation}
\gls{ReV}= \frac{\gls{V}\,\gls{c}}{\gls{nu}}\,,
\end{equation}
the normalised amplitude of motion, $\gls{Aoc}$, and the non-dimensional
stiffness of the spring
\begin{equation}
\gls{kn} = \frac{\gls{k}}{\gls{rhof}\gls{c;cb}\gls{V;sq}}\,.
\end{equation}
We use most of the governing parameters of \citet{arranz2022}, so that the 
present study also serves as reference for a tandem configuration studied 
with the same parameters \citep{schenk2026}.
The Reynolds number is $\gls{ReV}=\gls{c}\gls{V}/\gls{nu}=200$, the 
amplitude of the heaving motion is $\gls{A}/\gls{c}=0.5$ and the aspect 
ratio is $\gls{b}/\gls{c}=0.5$.
We consider three density ratios, $\gls{rhor}=1$, $10$ and $100$, which cover
the density-matched regime, the value used by \citet{arranz2022} and the
inertia-dominated regime.
For each density ratio, we carry out a parametric study of \gls{kn} to evaluate
its influence on propulsive performance and optimise its value to
maximise propulsive speed.
% -----------------------------------------------------------------------------

% =============================================================================
%   COMPUTATIONAL DOMAIN 
% =============================================================================
\subsection{Computational set-up}
\label{sec:meth:comp}
We perform \glspl{dns} with the in-house code TXPS,  which integrates the
Navier--Stokes equations for incompressible flow shown in \eqref{eq:ns},
together with equations \eqref{eq:lin} and \eqref{eq:mom} (in addition to the
above-mentioned constraints).
The flow solver employs second-order finite differences on a staggered grid and
a semi-implicit scheme (Adams--Bashforth/Crank--Nicolson), in which linear
terms are treated implicitly and nonlinear terms explicitly.
% -----------------------------------------------------------------------------

% -----------------------------------------------------------------------------
Following \citet{moriche2026}, we integrate the governing equations in a moving 
frame, an approach well suited for problems in which the motion of the bodies 
in the non-homogeneous direction is not known a priori.
It avoids excessively large domains in the streamwise (non-periodic) direction,
which would otherwise be needed to accommodate the unknown propulsive speed.
In the algorithm presented in \citet{moriche2026}, the settling velocity and the
Galileo number are replaced here by the propulsive speed and the Reynolds number
\gls{ReV}, respectively.
As indicated in \citet{moriche2026} we define two reference frames: i) the
laboratory frame, in which the fluid is at rest in the absence of flapper
motion, and ii) a moving frame translating with velocity $-U\gls{ex}$ with
respect to the laboratory frame.
The value of $U$ is obtained from an iterative procedure and, upon convergence,
matches (up to a given tolerance) the mean propulsive speed of the flapper
relative to the fluid, which is by definition the mean propulsive speed in the
laboratory frame.
As the moving frame is introduced only for computational purposes, all results 
are presented in the laboratory frame, which corresponds to a flapper 
self-propelling in quiescent fluid.
% -----------------------------------------------------------------------------

% -----------------------------------------------------------------------------
The computational domain is a cuboid with side lengths $[L_x,L_y,L_z]$, as
sketched in Figure~\ref{fig:meth:sketch}.
In the streamwise direction (\gls{vx_x}), we impose $\gls{vu}=U\gls{ex}$ at
the inlet and a convective outflow condition at the outlet.
The lateral and vertical directions (\gls{vx_y} and \gls{vx_z}) are periodic.
The leading edge of the flapper is initially placed $5\gls{c}$ downstream of
the inlet, and we require this distance to remain between $5\gls{c}$ and
$6\gls{c}$ for convergence, ensuring the (small) influence of the inlet is the
same in all cases.
With $L_x=16\gls{c}$, the wake can develop over $9\gls{c}$ to $10\gls{c}$
behind the trailing edge before it reaches the outflow boundary.
In the lateral and vertical directions, we choose $L_y=L_z=8\gls{c}$, which
is larger than the lateral extent of the domain used by \citet{arranz2022}.
A grid refinement study (Appendix~\ref{sec:app:gridstudy}) shows that a
resolution of $\gls{c}/\Delta x=64$ provides sufficient accuracy.
The resulting grid has $[1024 \times 512 \times 512]$ points.
The time step is $\Delta t=\gls{T}/2048$, where $\gls{T}=1/\gls{f}$ is the
period of the heaving motion, which keeps $CFL<0.3$ in all simulations.
% -----------------------------------------------------------------------------

% -----------------------------------------------------------------------------
The presence of the body is modelled using the \gls{ibm} originally proposed by 
\citet{uhlmann2005} and later extended by \citet{moriche2021_settling} to 
handle non-spherical bodies.
Since this formulation is unstable for density ratios approaching unity, the 
density-matched case $\gls{rhor}=1$ is computed with the \gls{ibm} of
\citet{schenk2025}.
\citet{schenk2025} compared both formulations and obtained good agreement for 
$\gls{rhor}\geq1.5$.
For $\gls{rhor}=10$ and $100$, we use the formulation of
\citet{moriche2021_settling}, which is about $7\%$ faster for the present
configuration.
Both formulations use the same distribution of $13\,952$ Lagrangian markers, 
which fill the volume of the flapper at the resolution $\gls{c}/\Delta x=64$.
Furthermore, here we incorporate ad hoc the term corresponding to the elastic
response of the torsional spring as shown in equation~\eqref{eq:meth:K}.
% -----------------------------------------------------------------------------

% =============================================================================
%   HYDRODYNAMIC/PERFORMANCE COEFFICIENTS
% =============================================================================
\subsection{Definitions to evaluate the hydrodynamic performance}
\label{sec:meth:performance}

% -----------------------------------------------------------------------------
In this section we present the performance indicators used to analyse the 
results presented below, adopting the definitions of~\cite{arranz2022}.
In all simulations, the flapper reaches a fully periodic state whose period
equals that of the imposed vertical motion, \gls{T}.
Since every oscillation cycle in this state is identical, we evaluate all mean
quantities by averaging over one cycle.
We start with the cycle-averaged propulsive speed
\begin{equation}\label{eq:meth:velocity}
    \gls{Up} = \frac{-1}{\gls{T}}\int_{t^*}^{t^*+\gls{T}} \gls{vup_x}(t)\,\wrt{t}\,,
\end{equation}
where $t^*$ denotes an arbitrary time instant within the periodic state and the 
minus sign ensures positive values of \gls{Up} for a flapper in self-propulsion
(swimming in the negative \gls{vx_x} direction).
To evaluate the mean input power of the flapper, we discard the negative values 
of the instantaneous input power 
\begin{equation} \label{eq:meth:IP}
    \gls{P} = \frac{1}{\gls{T}}\int_{t^*}^{t^*+\gls{T}} 
    \max(\gls{P_i}(t),\,0)\,\wrt{t}\,,
\end{equation}
where $\gls{P_i}=\gls{R}\cdot\gls{vup_z}$ and $\gls{R}$ is the vertical component 
of the actuator force at the centre of mass of the flapper, which follows from 
the \gls{vx_z}-component of the linear momentum equation
\begin{equation}
    \gls{R} = \gls{rhop}\gls{Vp} \ddt{\gls{vup_z}} - \gls{vF_z} \,.
  \label{eq:meth:actuator}
\end{equation}
The rationale behind neglecting negative values of \gls{P_i} to compute the 
mean input power is that we assume that the actuator has no means of recovering
energy, which is consistent with previous studies~\citep{berman2007,michelin2009}.
The torsional spring does store and return energy, but only in the
rotational degree of freedom, and in a passive manner; the present measure
concerns the active heaving actuation, for which no such recovery is assumed.
We also define the propulsive efficiency of a flapper as the ratio between the 
kinetic energy in the streamwise direction and the input work of one cycle
\begin{equation}
   \gls{E} = \frac{\frac{1}{2}\gls{rhop}\gls{Vp}\gls{Up;sq}}
    {\gls{P} \gls{T}} \,.
    \label{eq:meth:efficiency}
\end{equation}
This work-based definition is suited for problems in which the thrust/drag
force is not available \citep[for more details see][]{zhang2022}.
Please note that \gls{E} accounts for the kinetic energy of the flapper obtained
by self-propulsion, therefore it is proportional to the mass of the flapper and
makes the comparison between different density ratios difficult
\citep{arora2018}.
In order to further evaluate the propulsive efficiency of the flapper, we
complement the definitions of \citet{arranz2022} with a mass-specific quantity,
the non-dimensional cruising performance
\begin{equation}
    \gls{ve} = \frac{\gls{Up}/\gls{V}}
    {\gls{P}/ (\gls{rhof} \gls{V;cb} \gls{c;sq})} =  
    \gls{rhof} \gls{V;sq} \gls{c;sq}\frac{\gls{Up}}{\gls{P}}\,,
    \label{eq:meth:cruising}
\end{equation}
which measures the distance travelled per unit input work and has been used
by \citet{yeh2014,yeh2016}, under the name effectiveness, and by
\citet{arora2018}.
% -----------------------------------------------------------------------------

%% file: tex/03_results.tex
% =============================================================================
% =============================================================================
%   RESULTS
% =============================================================================
% =============================================================================
\section{Results} 
\label{sec:results}

% =============================================================================
%   EFFECT OF TORISONAL SPRING ON PROPULSION 
% =============================================================================
\subsection{Effect of the torsional spring stiffness on propulsion}
\label{sec:results:kinematics}

\begin{figure}[b!]
    \centering
    \includegraphics[width=0.8\textwidth]{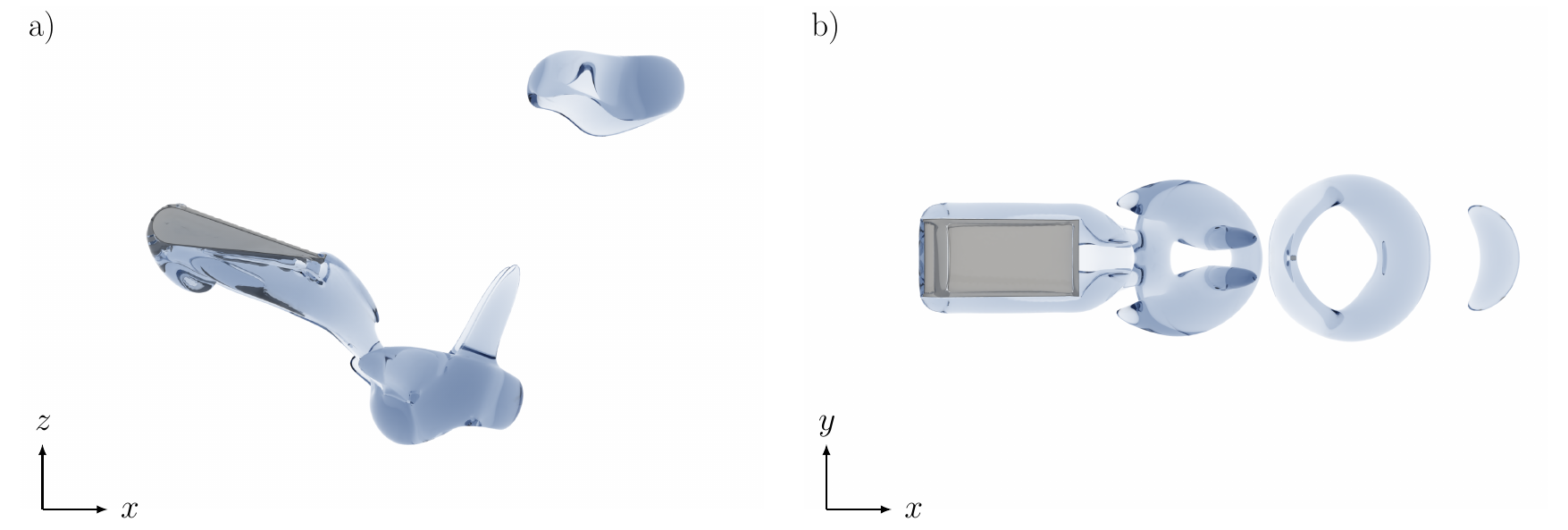}
    \caption{%
    a) Side and b) top views of the flow around the flapper with density ratio
    $\gls{rhor}=1$ and spring stiffness $\gls{kn}=\gls{kn1}$ at the end of the
    upstroke ($t/\gls{T} = 0.8125$).
    The flow is represented with iso-surfaces of the \gls{Q}-criterion
    ($\gls{Q} \gls{c;sq}/\gls{V;sq} = 0.5$).
    }
\label{fig:res:flapper_Q}
\end{figure}

% -----------------------------------------------------------------------------
We now proceed to analyse in detail the influence of the torsional spring
stiffness on the propulsive performance for one of the density ratios,
$\gls{rhor}=1$.
First, let us give an overview of the flow for the case in which the stiffness
of the spring is $\gls{kn}=\gls{kn1}$, where \gls{kn1} is the value that
maximises the propulsive speed (more details below).
Figure~\ref{fig:res:flapper_Q} shows iso-surfaces of the \gls{Q}-criterion
close to the end of the upstroke ($t/\gls{T} = 0.8125$).
At every stroke reversal ($t/\gls{T}=0,\,0.5$), a vortex ring  is shed into
the wake.
The trajectories of the rings are tilted with respect to the streamwise 
direction because of the effect of their self-induced velocity.
Rings shed at consecutive stroke reversals move in opposite directions, one
upwards and the other downwards.
As a result, the wake diverges into two branches of alternating circulation
vortex rings \citep{dong2006,kurt2020}.
% -----------------------------------------------------------------------------

% -----------------------------------------------------------------------------
%   Propulsive speed vs stiffness
% -----------------------------------------------------------------------------
\begin{figure}[t!]
    \centering
    \includegraphics[width=0.65\textwidth]{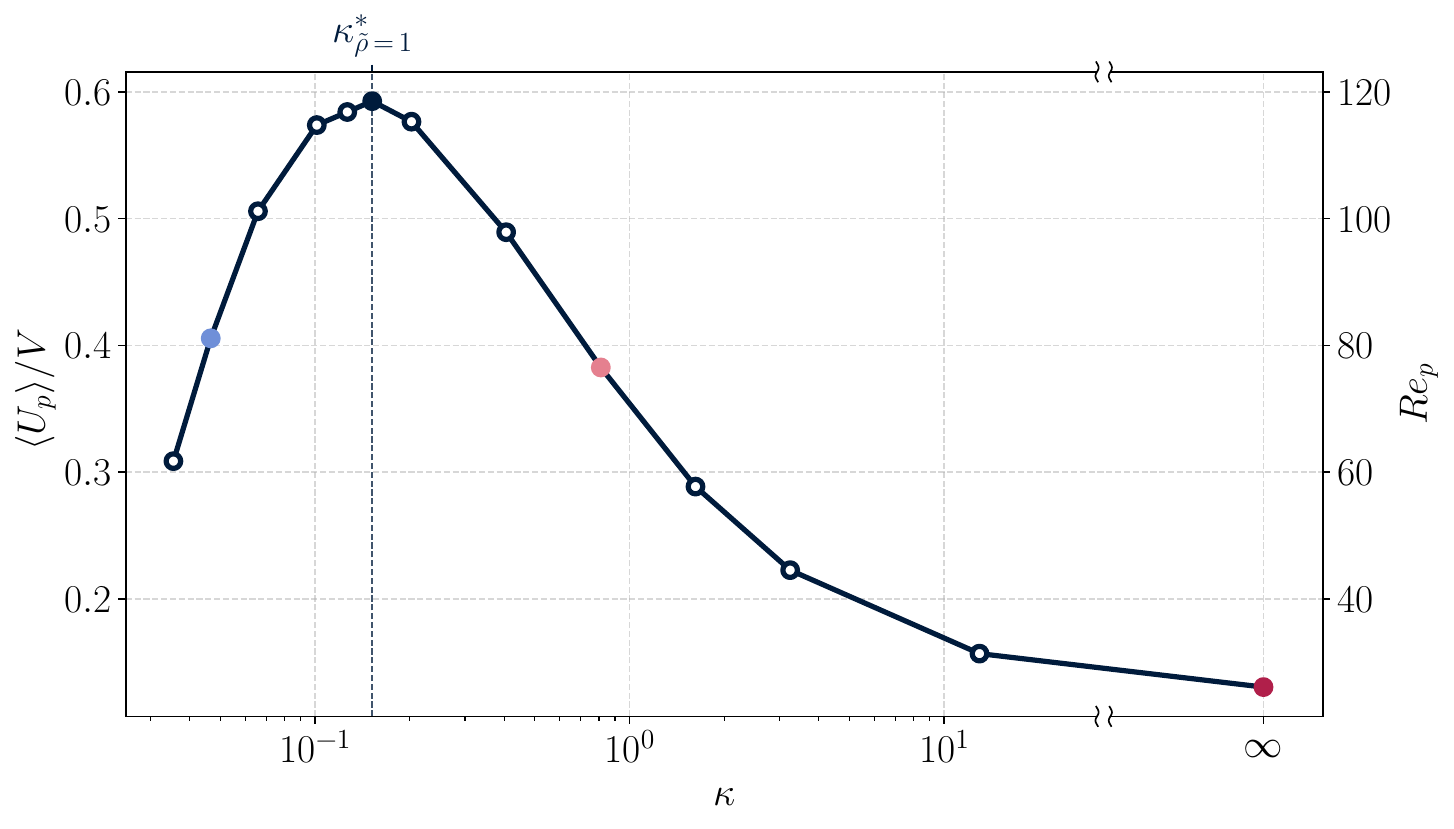}
    \caption{%
    Cycle-averaged propulsive speed $\gls{Up}/\gls{V}$ and propulsive
    Reynolds number \gls{Rep}, as a function of the
    normalised torsional spring stiffness \gls{kn}.
    The vertical dashed line indicates the spring stiffness that maximises
    propulsive speed, \gls{kn1}.
    Filled symbols highlight the four representative cases $\gls{kn} = 0.047$
    (soft), $0.15$ (optimum), $0.81$ (stiff) and $\gls{kn} = \infty$ (rigid).
    \label{fig:res:flapper}
    }
\end{figure}

% -----------------------------------------------------------------------------
To find the value of \gls{kn} that maximises the propulsive speed, we sweep
the values of the normalised spring stiffness in the range $\gls{kn}\in
[0.035,13]$, with the addition of the rigid flapper in which the pitching
motion is inhibited ($\gls{kn}=\infty$).
Figure~\ref{fig:res:flapper} shows the mean propulsive speed \gls{Up} as a
function of \gls{kn}, where a single optimum is clearly identified.
With the initially seeded values this optimum was found in the range
$\gls{kn}\in[0.1,0.2]$, thus we further refined this subrange to finally obtain
$\gls{kn}=0.15$ as the value that maximises propulsive speed.
We refer to this value as \gls{kn1}.
To facilitate the analysis of the effect of the stiffness, we select four
representative cases.
The first two are the case with infinite spring stiffness, labelled as
``rigid'', and the case with a spring stiffness that maximises propulsive
speed, denoted as ``optimum''.
The other two deliver similar propulsive speed ($\gls{Up}/\gls{V}\approx0.4$),
with a spring stiffness lower and higher than the optimum, and we label them
as ``soft'' and ``stiff'', respectively.
Please note that these cases have drastically different stiffness: there is
a factor of about $17$ between their respective values of \gls{kn}.
% -----------------------------------------------------------------------------

% -----------------------------------------------------------------------------
The current system shows self-propulsion for the upper limit of \gls{kn}
evaluated, which is the rigid case ($\gls{kn}=\infty$), reaching a propulsive
speed of $\gls{Up}/\gls{V}\approx0.13$.
Softening the spring from this limit gradually increases the propulsive
speed until the maximum of $\gls{Up}/\gls{V}=0.59$ is found (optimum case).
Reducing the stiffness further causes a sharp drop in the propulsive speed
(please note the logarithmic representation in Figure~\ref{fig:res:flapper}),
resulting in the loss of self-propulsion; the threshold value below which
the flapper does not self-propel was not determined.
The propulsive speed obtained with the optimum configuration corresponds
to a Reynolds number based on the propulsive speed  of 
$\gls{Rep}=\gls{Up}\gls{c}/\gls{nu}=118$.
For comparison, \citet{arranz2022} reported a maximum propulsive speed of
$\gls{Up}/\gls{V}=0.88$ ($\gls{Rep}=176$) for a chordwise-flexible plate
with the same kinematic parameters.
We attribute this difference to the distinct approaches used to model
the elastic response of the flapper to the hydrodynamic forces
(see \S~\ref{sec:meth:prob}).
If we compare the propulsive speed of the optimum case 
($\gls{kn}=\gls{kn1}$) with the rigid case ($\gls{kn}=\infty$), we observe an 
increase of  $350\%$.
This enhancement in \gls{Up} is notably larger than the one reported by
\citet{zhang2010} in a similar analysis in two dimensions.
For combined heaving and pitching motion, the Strouhal number can be computed
using the peak-to-peak excursion of the trailing edge instead of the amplitude
of the heaving motion \citep{smits2019}.
For the optimum case, we obtain an excursion of $\gls{ATE}=1.24\gls{c}$, and
the Strouhal number is then $\gls{St}=\gls{f}\gls{ATE}/\gls{Up}=0.67$, which
lies above the range $0.2\le\gls{St}\le0.4$ of cruise conditions found in 
swimming and flying animals \citep{taylor2003}.
A higher Strouhal number might be expected at the present low Reynolds number,
$\gls{Rep} = 118$, because the larger drag at low Reynolds numbers shifts
the preferred Strouhal number of swimmers to higher values
\citep{saadat2017,smits2019}.
At low Reynolds numbers, a pitching airfoil at $\gls{St}\approx0.3$ 
produces more drag than thrust \citep{smits2019}.
% -----------------------------------------------------------------------------

% -----------------------------------------------------------------------------
%   Passive pithcing, eff. AoA and CL
% -----------------------------------------------------------------------------
\subsubsection{Analysis of the passive pitching motion}
\begin{figure}[t!]
    \centering
    \includegraphics[width=0.9\textwidth]{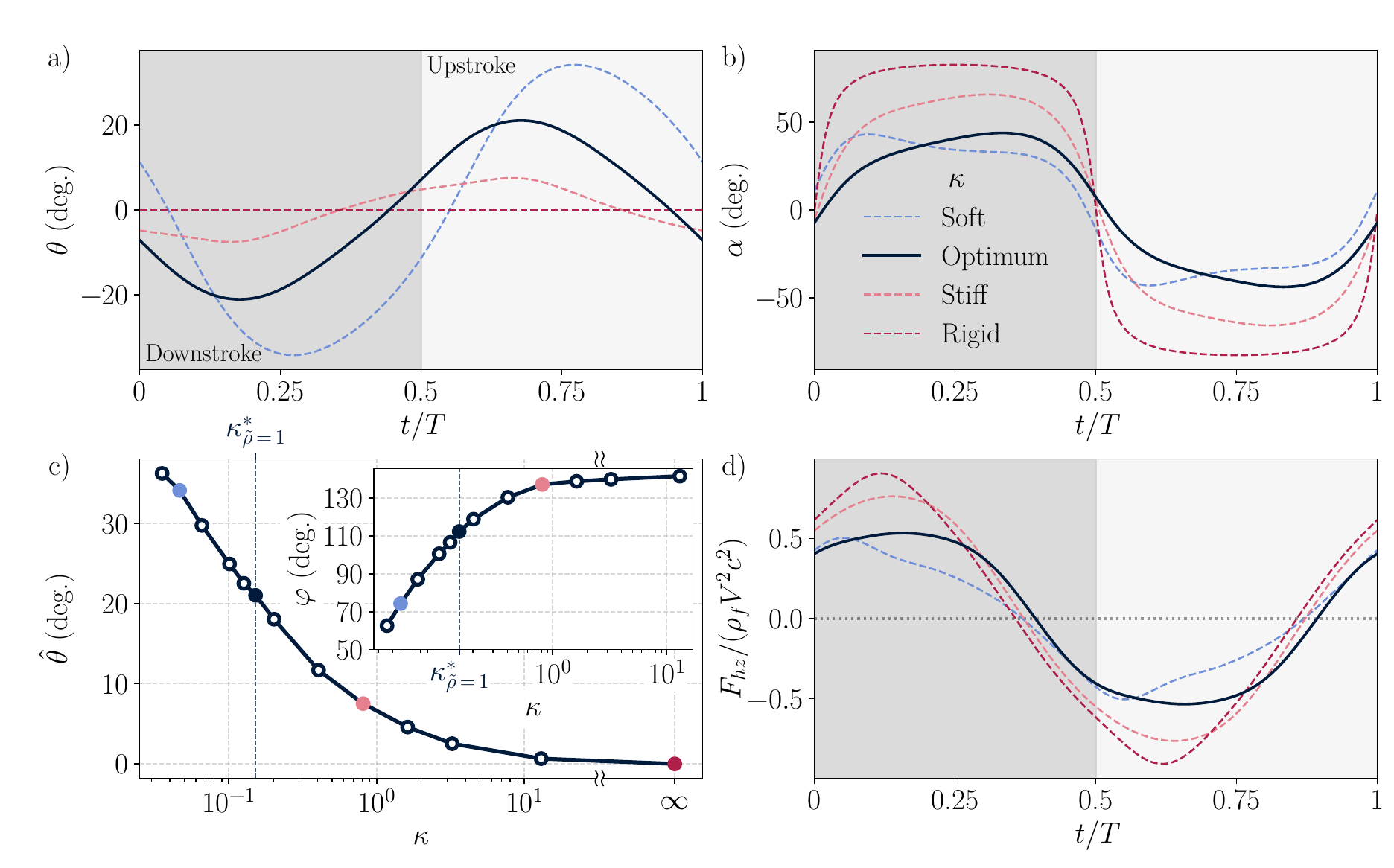}
    \caption{%
    a) Pitching angle $\gls{th}$ and b) effective angle of attack
    $\gls{ae}$ over one cycle for the selected cases.
    c) First Fourier mode of the pitching motion, \gls{thA}, as a function of
    \gls{kn}, with the inset showing the phase shift \gls{varphi} between the
    heaving motion and this mode.
    d) Vertical component of the hydrodynamic force $\gls{vF_z}/(\gls{rhof} \gls{V;sq} \gls{c;sq})$
    over one cycle for the representative stiffnesses.
    The shaded regions in a), b) and d) indicate the downstroke and upstroke, as
    labelled in a).
    The legend in b) applies to panels a), b) and d), and the dark-blue
    vertical dashed lines in c) mark $\gls{kn1}$.
    }
\label{fig:res:single:quat}
\end{figure}

% -----------------------------------------------------------------------------
We now turn our attention to the passive pitching motion, which results from 
the balance between the hydrodynamic moment, the rotational inertia of the 
flapper and the spring response.
Figure~\ref{fig:res:single:quat}a) shows the instantaneous evolution of the
pitching angle \gls{th} over one cycle for the selected cases.
The pitching angle for the rigid case is $\gls{th}=0$ and it is included 
for completeness.
The modulus of the heaving velocity (common to all the cases) is largest
at the middle of both downstroke ($t/\gls{T}=0.25$) and upstroke
($t/\gls{T}=0.75$), and vanishes at stroke reversals ($t/\gls{T}=0$ and $0.5$).
To support the following discussion, let us define as the reference feathering
motion the one in which the flapper is aligned with the horizontal
($\gls{th}=0$) at the stroke reversals, and pitch reversal (largest pitching
angle) takes place at the mid-strokes.
The phase shift between this reference feathering motion and the imposed heaving
would be $90^\circ$.
We observe a predominantly sinusoidal behaviour of \gls{th} in all cases
evaluated, however higher-order harmonics are visible in both the soft and stiff
cases, showing the non-trivial influence of the hydrodynamic forces on the
flapper's response.
For the optimum case ($\gls{kn}=\gls{kn1}$), \gls{th} peaks before
both mid-strokes, which implies that its angular velocity is non-zero when 
the vertical motion reaches its maximum speed and at stroke reversals.
It passes through $\gls{th}=0$ shortly before the end of each half stroke,
so the foil is horizontally aligned preceding stroke reversal, i.e., the
pitching motion leads the reference feathering behaviour.
For the soft case, the largest pitching angle occurs slightly after both
mid-strokes, and $\gls{th}=0$ takes place slightly after the end of 
both strokes, therefore softening the stiffness results in a lagged pitching
motion with respect to the optimum case, and also the reference feathering
behaviour.
The amplitude of the pitching motion of the soft case is notably larger than
the one obtained for the optimum case.
A similar dependence was reported by \citet{huang2023} for passively pitching
wings in hovering flight, where the performance drops once the foil is
horizontally aligned after stroke reversal.
The stiff case shows the opposite behaviour: the resultant pitching motion
leads that of the optimum case and its amplitude is notably smaller.
This is to be expected because the softer the spring, the less resistance
is found to counteract the hydrodynamic moment, and vice versa for the stiff
case.
% -----------------------------------------------------------------------------

% -----------------------------------------------------------------------------
%
We proceed now to characterise the pitching motion and its relation with the
imposed heaving.
Figure~\ref{fig:res:single:quat}c) shows the amplitude of pitching motion,
\gls{thA}, which is defined as the amplitude of its first Fourier mode.
We use this definition because for some cases, typically the ones with
the stiffest and the softest springs, the pitching motion shows 
higher-order harmonics.
This is particularly important to perform a consistent analysis when
characterising the phase shift between heaving and pitching motions.
\gls{thA} decreases monotonically with \gls{kn} for the cases evaluated,
from a value of $\gls{thA}=36.3^\circ$ for the softest spring to
$\gls{thA}\approx 0$ for the stiffest non-rigid spring.
The same trend was reported for passively pitching wings with a
torsional spring \citep{spagnolie2010,xiao2014}.
Interestingly, the optimum does not correspond to the largest pitching
amplitude but to an intermediate value of $\gls{thA}=21.1^\circ$, which
is consistent with the findings of \citet{zhang2010} for a two-dimensional
plate with a torsional spring at its leading edge, where the largest forward
speeds occur for intermediate pitching amplitudes between $5^\circ$ and
$25^\circ$.
Besides the amplitude, the stiffness also influences the phase shift
between heaving and pitching motion, \gls{varphi}.
Figure~\ref{fig:res:single:quat}c) shows that the phase shift \gls{varphi}
increases monotonically with \gls{kn} for the cases evaluated.
Cases in which $\gls{kn}<0.05$ show a pitching motion that lags the reference
feathering behaviour, whereas cases with $\gls{kn}>0.05$ lead.
This is indeed illustrated in the selected soft and stiff cases, the former
resulting in $\gls{varphi}=74^\circ < 90^\circ$ and the latter in 
$\gls{varphi}=142^\circ>90^\circ$.
The phase shift of the optimum case ($\gls{varphi}=113^\circ$) is close to
the one reported by \citet{moriche2017}, who found the maximum mean thrust
coefficient for a two-dimensional airfoil with prescribed heaving and pitching
motions in a free stream takes place when the phase shift between the two 
motions is $110^\circ$.
This suggests that, at the optimum, the torsional spring passively sets a
phase shift close to the one that maximises thrust for prescribed kinematics.
% -----------------------------------------------------------------------------

% -----------------------------------------------------------------------------
%   Effective angle of attack
% -----------------------------------------------------------------------------
Now, let us focus on the effective angle of attack, \gls{ae}.
The effective angle of attack differs from the pitching angle because its
definition includes the inclination of the free stream seen by the airfoil
due to its vertical motion.
The value of \gls{ae} is directly linked with the magnitude of the hydrodynamic
forces originated by circulatory effects, and it is of central importance in
self-propulsion (even a non-pitching foil can generate thrust because of the
generation of a non-zero effective angle of attack due to its vertical 
oscillation).
The effective angle of attack is given by
\begin{equation}\label{eq:ae}
    \gls{ae} = \gls{th} - \arctan\left(\frac{\gls{vup_z}}{\gls{Up}}\right)
             = \gls{th} + \arctan\left(\frac{\gls{V}\sin(2\pi\gls{f}t)}
                                            {\gls{Up}}\right) \,.
\end{equation}
The contribution from the vertical motion (second term) is positive during the
downstroke, and negative during the upstroke.
Figure~\ref{fig:res:single:quat}a) shows that the pitching angle is negative
during most of the downstroke, and positive during the upstroke, which is the
opposite behaviour of the term $-\arctan(\gls{vup_z}/\gls{Up})$.
Since the values of $-\arctan(\gls{vup_z}/\gls{Up})$ are larger in
magnitude than the ones obtained for \gls{th}, the effective angle of
attack turns out  to be positive (negative) in the downstroke (upstroke).
The pitching motion thus acts as a damper, or softening mechanism, for the
potentially large values that \gls{ae} would take in the absence of pitching.
For the rigid case, the propulsive speed is small compared to the maximum
heaving velocity, therefore \gls{ae} rises abruptly at the beginning of each 
half stroke and remains slightly above $80^\circ$ during most of it.
Such large values of \gls{ae} result in flow separation at both leading and 
trailing edges.
For the stiff case, \gls{ae} is approximately $50^\circ$ during most of
each stroke.
From the rigid case to the stiff case, both the pitching motion and the
increase in propulsive speed contribute to a reduction in the effective
angle of attack, however the latter is the dominant effect.
The optimum case shows a gradual increase of \gls{ae} during each stroke,
reaching its maximum value of about $44^\circ$ at 
$t/\gls{T}\approx0.33$, when the flapper is still tilted.
The positive contribution of this combination of tilting and large effective
angle of attack is discussed below.
For the soft case, \gls{ae} shows again an abrupt increase at the beginning
of each stroke, later remaining between about $30^\circ$ and $40^\circ$
during most of each stroke.
The vertical hydrodynamic force \gls{vF_z} in 
Figure~\ref{fig:res:single:quat}d) shows that the non-optimum cases, which
present an abrupt increase of \gls{ae} at the beginning of each stroke, show
a single peak in its evolution, followed by a drop in the force obtained.
It seems that the slower progression in \gls{ae} obtained in the optimum case
helps the force to build in a more stable manner, where the effective angle of
attack and the resultant force are tightly coupled.
There is a clear trend in the evolution of the maximum vertical hydrodynamic
force obtained for the rigid, stiff and optimum cases: the lower the effective
angle, the lower the force obtained, whereas the soft case presents values of
\gls{ae} similar to those of the optimum case, but a notably smaller force.
%-------------------------------------------------------------------------------

%-------------------------------------------------------------------------------
These results indicate that the stiffness of the flapper influences the
propulsive speed through two mechanisms: i) a purely geometrical effect in 
which large pitching amplitudes tilt the circulatory component of the 
aerodynamic force in the direction of motion, and ii) the effect of the 
effective angle of attack, which is directly related to the magnitude of 
the hydrodynamic force.
In order to understand better the interaction between these two mechanisms, let
us compare the four selected cases.
Figure~\ref{fig:res:single:quat}d) shows how the vertical force is maximum
for the rigid flapper, however, because of the absence of pitching
($\gls{th}=0$), there is no upstream tilting of the generated force.
The stiff case presents a small tilting and only slightly smaller vertical
forces.
For the soft case the flapper shows the largest tilting
(see Figure~\ref{fig:res:single:quat}a), but the modulus of the force
generated is clearly smaller and the propulsive speed drops.
This trade-off is clearest for the stiffer and softer springs, which reach
similar propulsive speeds: the stiffer spring keeps \gls{vF_z} large but
tilts by only about $7.5^\circ$, whereas the softer spring tilts by about
$34^\circ$ but reaches only $\gls{vF_z}\approx0.5$.
The optimum is thus a balance, in both amplitude and timing of the pitching
motion, between insufficient pitching in the stiff limit and excessive
pitching in the soft limit.
%-------------------------------------------------------------------------------

%-------------------------------------------------------------------------------
% FLOW VISUALISIATION
%-------------------------------------------------------------------------------
\subsubsection{Flow visualisation}
\begin{figure}[t!]
    \centering
    \includegraphics[width=0.75\textwidth]{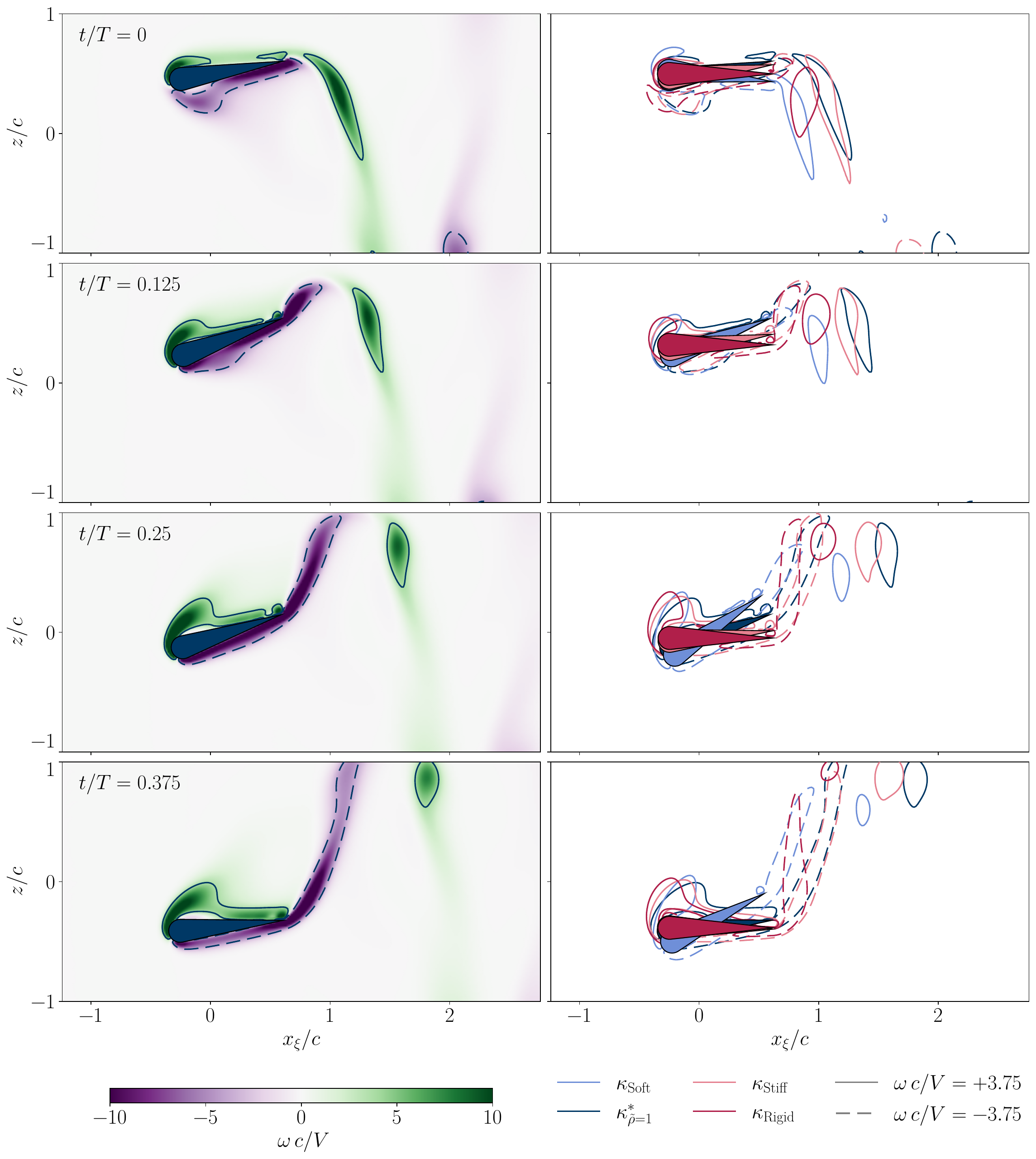}
    \caption{%
    Vorticity in the midplane around the flapper with $\gls{rhor}=1$ at four
    equispaced instants of the downstroke.
    The left column shows filled contours of the optimum case ($\gls{kn}=\gls{kn1}$)
    superimposed with line contours for $\gls{Vort} \, \gls{c}/\gls{V} = \pm 3.75$.
    The right column shows line contours with the same threshold as in the left column
    including the four representative cases.
    The cases are shifted in the streamwise direction by a constant value so that
    the mean positions of the centres of mass coincide; $x_\xi$ is then the streamwise
    coordinate shifted by the corresponding value.
    %coordinateis measured from the mean 
    %center of mass of $\kappa^*_{\tilde{\rho}=1}$.} 
    \label{fig:res:single:phases}
    }
\end{figure}
To illustrate the above-mentioned flow separation and its relation with
\gls{kn}, we compare spanwise vorticity fields ($\gls{Vort} = \partial
\gls{vu_x}/\partial \gls{vx_z} - \partial \gls{vu_z}/\partial \gls{vx_x} $)
at the midplane (\gls{vx_x}--\gls{vx_z}), for each of the selected cases.
In Figure~\ref{fig:res:single:phases} we display four equally spaced instants
of the downstroke, $t/\gls{T} = 0,0.125,0.25,0.375$; the upstroke is
mirror-symmetric and therefore not shown.
Filled contours of \gls{Vort} are used for the optimum case and contour lines 
of the four cases together to highlight the differences.
It can be seen how the leading-edge vortex formed for the optimum case
grows as the flapper heaves, remaining attached at the four time instants
represented.
On the lower surface, a shear layer is generated.
The intensity of the shear layer increases from $t/\gls{T}=0$ to $t/\gls{T}=
0.25$ and is continuously shed into the wake, to later roll up into the
trailing-edge vortex.
The roll up of the shear layer is visible from the shear layer shed into the
wake in the previous stroke (upstroke).
The trailing-edge vortex was shed during the previous half stroke and is
advected downstream with a vertical drift.
If we analyse the differences between the selected cases, the most striking
feature is that the leading-edge vortex for the optimum case shows high
compactness, and the largest extension  over the upper surface.
The largest detachment is seen for the rigid case, which is to be expected
because of the largest values of \gls{ae} reached in this case.
At the trailing edge, the shear layer of the rigid flapper leaves almost
perpendicular to the chord, whereas for lower \gls{kn} it leaves more aligned
with the chord.
Overall, the rigid flapper and the flapper with the stiffer spring, which have 
the largest \gls{ae}, show stronger flow separation at the leading  edge, and 
the rigid flapper also at the trailing edge.
The wake also reflects the different advection of the vortices, which is 
 related to the propulsive speed.
The trailing-edge vortex generated by the optimum case is the one found
furthest downstream, and the one generated by the rigid case is the closest
to the flapper at $t/\gls{T}=0.375$.
It is interesting to see how the vortices shed by the soft and stiff cases
present a somewhat different advection speed, although these two cases reach 
approximately the same propulsive speed.
% -----------------------------------------------------------------------------

% -----------------------------------------------------------------------------
% Power and efficiency 
% -----------------------------------------------------------------------------
\subsubsection{Power and efficiency}
\begin{figure}[t!]
    \centering
    \includegraphics[width=\textwidth]{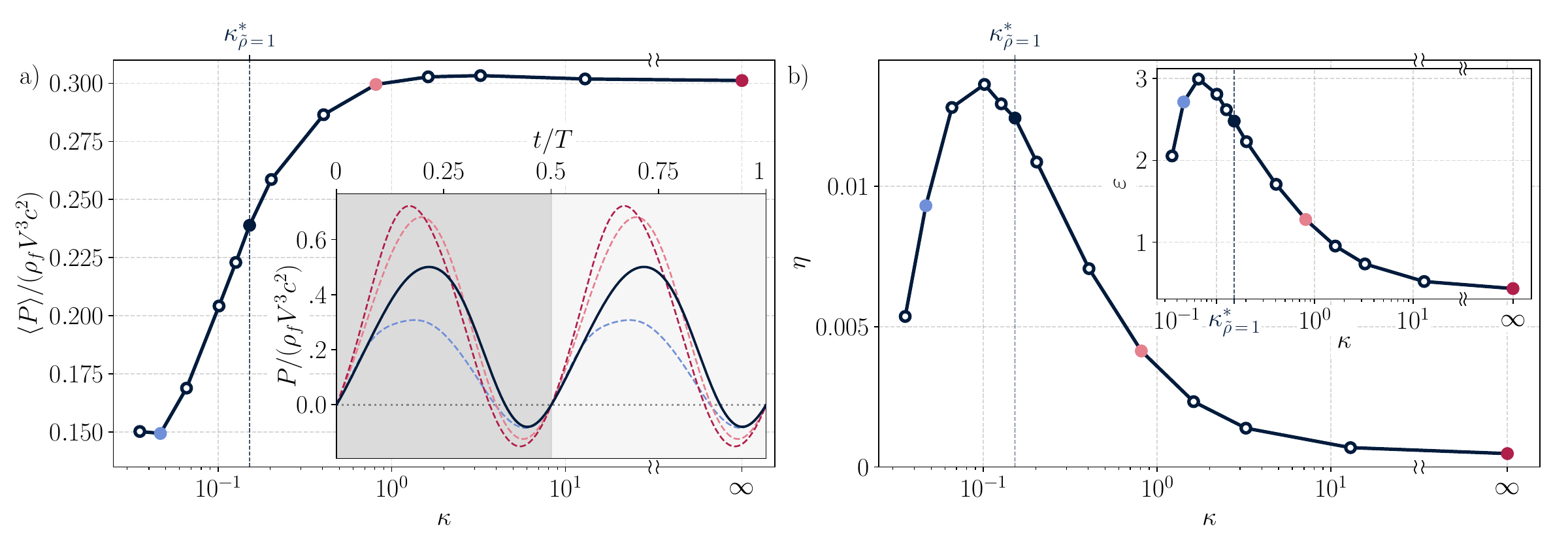}
    \caption{%
    a) Cycle-averaged input power $\gls{P}/ (\gls{rhof} \gls{V;cb} \gls{c;sq})$ 
    and
    b) propulsive efficiency $\gls{E}$ as functions of $\gls{kn}$.
    The inset in a) shows the input power $\gls{P_i}$ over one cycle for the
    representative stiffnesses, and the inset in b) the cruising performance
    \gls{ve} as a function of $\gls{kn}$.
    The dark and light shaded regions in the inset in a) mark the downstroke
    and upstroke, respectively.
    The legend in Figure~\ref{fig:res:single:quat}b) applies to the inset in 
    a), and the vertical dashed lines mark $\gls{kn1}$.
   }
\label{fig:res:single:poweta}
\end{figure}
We now consider the mean input power \gls{P}, the propulsive efficiency
\gls{E} and the cruising performance \gls{ve}, shown in
Figure~\ref{fig:res:single:poweta}.
The mean input power increases monotonically with the stiffness in the range
$\gls{kn}\in[0.05,2]$, taking values from $\gls{P}=0.15$ to $0.3$.
Reducing the stiffness below $\gls{kn}\approx 0.05$ results in a small increase
in \gls{P}.
This behaviour might seem spurious, but it is actually observed more consistently
for higher density ratios (see the inset in 
Figure~\ref{fig:res:flapper:opt:power}a) presented below).
Increasing the stiffness from $\gls{kn}\approx2$ results in a small,
almost negligible, reduction in \gls{P}, which we interpret as a plateau.
The increase in mean input power found in the range $\gls{kn}\in[0.05,2]$ is
accompanied by a reduction of the pitching amplitude \gls{thA}.
This is to be expected since stiffer springs inhibit pitching, thus the 
flapper meets the flow at a larger \gls{ae}, which is directly related with 
the vertical force (as shown in Figure~\ref{fig:res:single:quat}b,d).
For the softest spring ($\gls{kn}\approx 0.035$), however, \gls{P} is
slightly larger than for the soft case ($\gls{kn}\approx 0.047$).
For this spring, the propulsive speed is low (Figure~\ref{fig:res:flapper}), 
so the incoming flow is more strongly inclined with respect to the swimming 
direction.
In addition, the pitch angle is still positive at stroke reversal, therefore
it shows horizontal alignment ($\gls{th}=0$) only after the stroke reversal,
as shown for the soft case in Figure~\ref{fig:res:single:quat}a).
At the beginning of each half stroke, the flapper is thus still tilted as in
the previous half stroke, which results in a peak of \gls{ae} at the 
beginning of each half stroke.
This is already visible for the soft case in 
Figure~\ref{fig:res:single:quat}b), where \gls{ae} shows an additional peak
shortly after each stroke reversal.
For the softest spring, this peak is even larger (not shown).
The instantaneous evolution of the input power \gls{P_i} for the selected cases
is shown in the inset of Figure~\ref{fig:res:single:poweta}a).
In each half stroke \gls{P_i} is positive during its first part and becomes 
negative before the stroke reversal.
These negative values do not enter the cycle average \gls{P}
(equation~\eqref{eq:meth:IP}).
The positive peaks of \gls{P_i} follow the same trend as \gls{vF_z} with 
\gls{kn}, decreasing from about $0.72$ for the rigid flapper to about $0.3$ 
for the soft case.
This is consistent with the larger vertical force of the stiffer cases
requiring more input power.
% -----------------------------------------------------------------------------

% -----------------------------------------------------------------------------
Now, let us focus on the efficiency \gls{E} and cruising performance \gls{ve}.
When comparing the input power of the rigid and stiff cases, these are almost 
the same, however the propulsive speed of the stiff case is nearly three times 
larger than the one of the rigid case (Figure~\ref{fig:res:flapper}).
Thus, a small pitching amplitude increases the propulsive speed at almost no
additional input power.
For the optimum case, the propulsive speed is $4.5$ times larger than that of 
the rigid case, and the input power is reduced by a factor of 
approximately $0.8$.
Since \gls{E} scales with $\gls{Up;sq}/\gls{P}$, \gls{E} is about
$4.5^2/0.8\approx25$ times larger.
It is interesting to note that the optimum of \gls{kn} that maximises
propulsive speed does not coincide with the value that maximises propulsive
efficiency.
A maximum of \gls{E} of about $0.014$ is obtained for $\gls{kn}\approx0.10$,
which is slightly smaller than \gls{kn1}, whereas the optimum case
($\gls{kn}=\gls{kn1}$) reaches $\gls{E}\approx0.013$.
This is because around these values of \gls{kn} the propulsive speed barely
changes while the input power shows a clear drop with decreasing \gls{kn}.
As \gls{kn} decreases even further, the propulsive speed falls sharply, and
the lower input power is not enough to prevent \gls{E} from falling too.
A similar difference between the optima was reported by \citet{xiao2014} for
a three-dimensional wing with an elliptical cross-section and a torsional
spring, where the thrust is largest for the softest spring considered while
the efficiency peaks at an intermediate stiffness.
For flexible plates, in contrast, \citet{tang2016} and \citet{wang2020}
found the maximum propulsive speed and efficiency at nearly the same
stiffness.
The inset of Figure~\ref{fig:res:single:poweta}b) shows the cruising
performance \gls{ve}, i.e., the distance travelled per unit input work
\citep{yeh2014,yeh2016,arora2018}.
For the optimum case ($\gls{kn}=\gls{kn1}$) we obtain $\gls{ve}=2.5$, which
is $5.6$ times larger than that of the rigid flapper, which follows the same
rationale explained above: an enhancement of $4.5$ times the propulsive speed
and a reduction of the input power by a factor of $0.8$.
The maximum of $\gls{ve}\approx3.0$ lies at an even softer spring than
that of \gls{E}, at $\gls{kn}\approx0.066$, since \gls{ve} contains
\gls{Up} only linearly, whereas \gls{E} scales with $\gls{Up;sq}/\gls{P}$.
\citet{yeh2014} also found the largest distance travelled per unit input
work for a three-dimensional elastic plate at a lower propulsive speed 
than the maximum one.
For softer springs, \gls{ve} drops sharply, following the drop of the
propulsive speed.
% -----------------------------------------------------------------------------

% =============================================================================
%   TRIOLOGY
% =============================================================================
\subsection{Density and torsional spring stiffness: a common flow state} 
\label{sec:results:optimum}

% -----------------------------------------------------------------------------
%   III - Proppulsive speed
% -----------------------------------------------------------------------------
\begin{figure}[t!]
    \centering
    \includegraphics[width=0.7\textwidth]{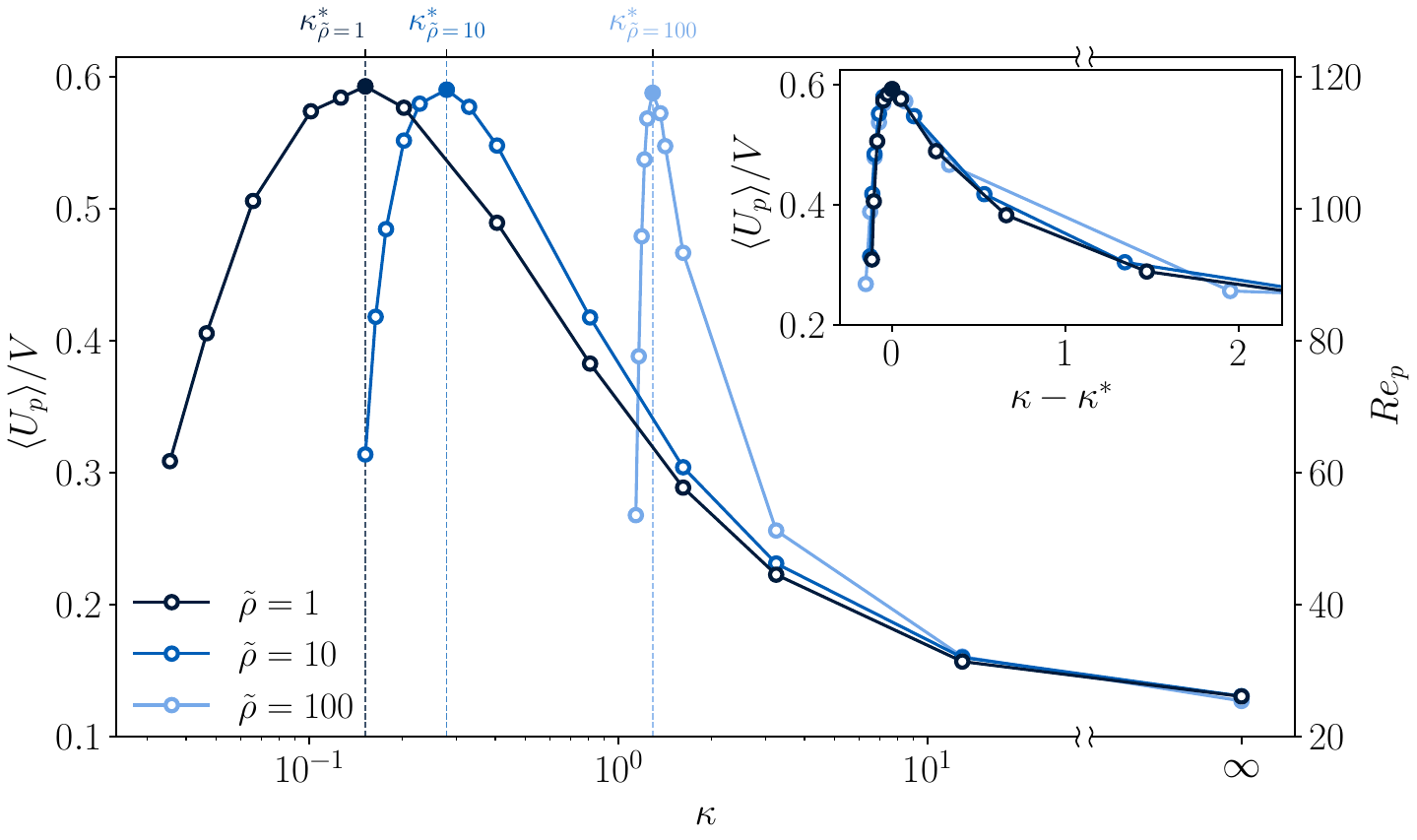}
    \caption{%
    Cycle-averaged propulsive speed $\gls{Up}/\gls{V}$ as a function of the
    torsional spring stiffness $\gls{kn}$ for the density ratios
    $\gls{rhor} = 1$, $10$ and $100$.
    The right axis gives the corresponding propulsive Reynolds number
    $\gls{Rep}$.
    Filled symbols mark the optimal stiffnesses $\gls{kn1}$, $\gls{kn10}$ and
    $\gls{kn100}$, which are also indicated by the vertical dashed lines.
    The inset shows the same data as a function of the stiffness relative to
    the respective optimum, $\gls{kn} - \gls{knOPT}$, on a linear scale.
    The break in the abscissa indicates that the rigid case is not drawn to
    scale.}
    \label{fig:res:opt:Up}
\end{figure}
We now extend the analysis to the three density ratios $\gls{rhor}=1$, $10$
and $100$.
For each density ratio, we repeat the optimisation procedure for \gls{kn} 
to maximise propulsive speed described in \S~\ref{sec:results:kinematics}
(see Figure~\ref{fig:res:opt:Up}).
The optimum values of \gls{kn} obtained are $0.15$, $0.28$ and $1.29$ for
$\gls{rhor}=1$, $10$ and $100$, respectively, which we refer to as
\gls{kn1},  \gls{kn10} and  \gls{kn100}.
Hence, the higher the density ratio, the stiffer the torsional spring needed 
to reach the maximum propulsive speed.
Interestingly, the maximum propulsive speed obtained is almost the same for
all three density ratios, $\gls{Up}/\gls{V}=0.59$, which corresponds to
$\gls{Rep}=118$ and $\gls{St}=0.67$.
Starting from the rigid flapper, all three density ratios self-propel at the
same speed, $\gls{Up}/\gls{V}\approx0.13$.
The three curves still coincide down to the stiffest spring,
$\gls{kn}=13$.
As we soften the spring, the propulsive speed increases gradually for all
three density ratios.
The curves separate, and the higher the density ratio, the stiffer the 
spring at which this increase sets in.
Each curve reaches its maximum at the respective optimum, and softening the
spring further causes a sharp drop in propulsive speed, as already observed
for $\gls{rhor}=1$ in \S~\ref{sec:results:kinematics}.
The inset of Figure~\ref{fig:res:opt:Up} shows the same data as a function
of $\gls{kn}-\gls{knOPT}$ on a linear scale.
Shifted in this way, the curves of the three density ratios collapse onto a
single curve.
This representation also highlights the sharp drop of the propulsive velocity
for stiffnesses below the optimum.
% -----------------------------------------------------------------------------

% -----------------------------------------------------------------------------
%   Resonance
% -----------------------------------------------------------------------------
To test whether the optimum corresponds to a resonance of the passive
pitching motion, we compare the stiffness that maximises the propulsive
speed with the resonant stiffness.
Since we perform this analysis for each of the density ratios, we use the
symbol \gls{knOPT} to identify the optimum, independently of its density 
ratio.
In quiescent fluid, the flapper oscillates in pitch with the natural
frequency $f_n=\sqrt{\gls{k}/(I_p+I_a)}/(2\pi)$.
Here, \gls{k} is the dimensional spring stiffness, 
$I_p=\gls{rhop}\gls{Vp}\gls{J}$ the moment of inertia of the flapper 
about its centre of mass (Appendix~\ref{sec:app:flapper_geometry}) 
and $I_a$ the added moment of inertia of the fluid about the same axis.
We estimate $I_a$ for a thin plate in potential flow, using the added-mass
coefficients of \citet{brennen1982} with a correction for the finite span
(Appendix~\ref{sec:app:added_inertia}).
Since we estimate $I_a$ for a thin plate, it does not account for the finite
thickness and the rounded leading edge of the flapper.
Furthermore, the potential flow estimate neglects viscosity and the mean
flow, and assumes small pitching amplitudes.
The resonance condition $f_n=\gls{f}$ then gives
$\kappa_{res}=0.0117\,\gls{rhor}+0.047$, where the first term contains
the inertia of the flapper and the second the added inertia of the fluid.
For $\gls{rhor}=1$, $10$ and $100$, this yields $\kappa_{res}\approx0.058$, 
$0.16$ and $1.21$, whereas the optima lie at $0.15$, $0.28$ and $1.29$.
The optimal spring is therefore stiffer than the resonant spring for all
three density ratios, and the optimum does not correspond to resonance.
The difference between \gls{knOPT} and $\kappa_{res}$ is nearly the same for
all three density ratios, although \gls{knOPT} itself increases strongly
with \gls{rhor}.
This indicates that \gls{knOPT} grows with \gls{rhor} like the inertia of
the flapper, while the fluid adds a nearly constant contribution.
Previous studies also found the optimal spring to be stiffer than the
resonant one, both for the propulsive speed in two-dimensional simulations
\citep{zhang2010,arora2018} and for the thrust in a linear two-dimensional
theory \citep{moore2015}.
Experiments with flexible flapping wings show the same, with the largest
thrust power for wings stiffer than resonant \citep{ramananarivo2011}.
In contrast, \citet{fernandez2021} found the best performance close to the 
natural frequency of the coupled fluid--structure system, which includes the 
hydrodynamic contributions of the mean flow.
\citet{arora2018} further showed that the maximum thrust does not coincide
with the highest propulsive speed of a self-propelled plate.
This agrees with our results for $\gls{rhor}=1$ presented in
\S~\ref{sec:results:kinematics}: the rigid and stiff cases produce the 
largest hydrodynamic force, but the highest propulsive speed requires a 
softer spring that tilts this force towards the swimming direction.

% -----------------------------------------------------------------------------
%   III - Phase optimum
% -----------------------------------------------------------------------------
\begin{figure}[t]
    \centering
    \includegraphics[width=0.8 \textwidth]{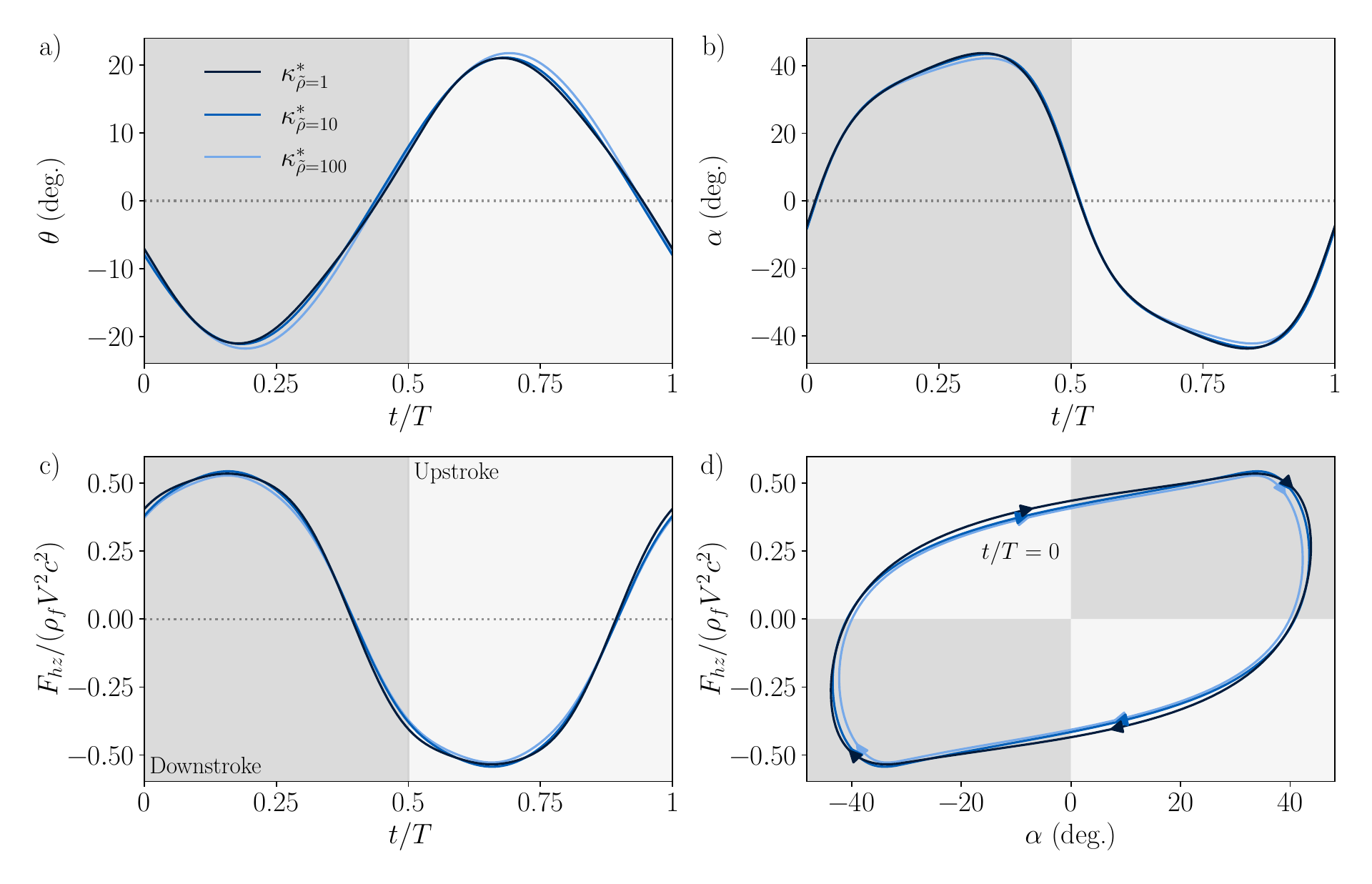}
    \caption{%
    a) Pitching angle $\gls{th}$, b) effective angle of attack $\gls{ae}$ and
    c) vertical hydrodynamic force $\gls{vF_z}/(\gls{rhof} \gls{V;sq} \gls{c;sq})$ 
    over one cycle, and d) vertical component of the hydrodynamic force 
    $\gls{vF_z}/(\gls{rhof} \gls{V;sq} \gls{c;sq})$ as a function of the 
    effective angle of attack $\gls{ae}$ during one period at the optimal 
    stiffnesses $\gls{kn1}$, $\gls{kn10}$ and $\gls{kn100}$.
    The shaded regions in a)--c) mark the downstroke and upstroke, as
    labelled in c).
    The arrows in d) mark $t/\gls{T} = 0$, $0.25$, $0.5$ and $0.75$ and
    indicate the direction of the loop.
    The legend in a) applies to all panels.}
    \label{fig:res:opt:phase}
\end{figure}

% -----------------------------------------------------------------------------
At their optimal stiffness, \gls{kn1}, \gls{kn10} and \gls{kn100}, the three
flappers reach the same propulsive speed, although the optimal stiffness
differs by a factor of about $8.5$ between $\gls{rhor}=1$ and $100$.
Their pitching angles nearly coincide, with an amplitude \gls{thA} of
$21.1^\circ$, slightly larger for $\gls{rhor}=100$, and the flapper is 
horizantlly aligned ($\gls{th}=0$) shortly before stroke reversal 
(Figure~\ref{fig:res:opt:phase}a).
Since \gls{ae} depends only on \gls{th}, the heaving velocity and \gls{Up},
the effective angle of attack is also the same, with a maximum of about
$44^\circ$ at $t/\gls{T}\approx0.33$ (Figure~\ref{fig:res:opt:phase}b).
Consequently, the vertical hydrodynamic forces collapse as well, with an
amplitude of about $0.54$ (Figure~\ref{fig:res:opt:phase}c).
Plotted as a function of \gls{ae}, \gls{vF_z} describes nearly identical loops
for the three flappers, which produce most of their force in the first half of
each half stroke (Figure~\ref{fig:res:opt:phase}d).
At the optimum, the three density ratios therefore reach essentially the same
kinematic and hydrodynamic state.
The density ratio only changes the stiffness needed to reach this state,
since a flapper of higher density ratio has a larger rotational inertia and
requires a stiffer spring to produce the same pitching motion.
% -----------------------------------------------------------------------------

% -----------------------------------------------------------------------------
%   III - Kinematics
% -----------------------------------------------------------------------------
\begin{figure}[t]
    \centering
    \includegraphics[width=\textwidth]{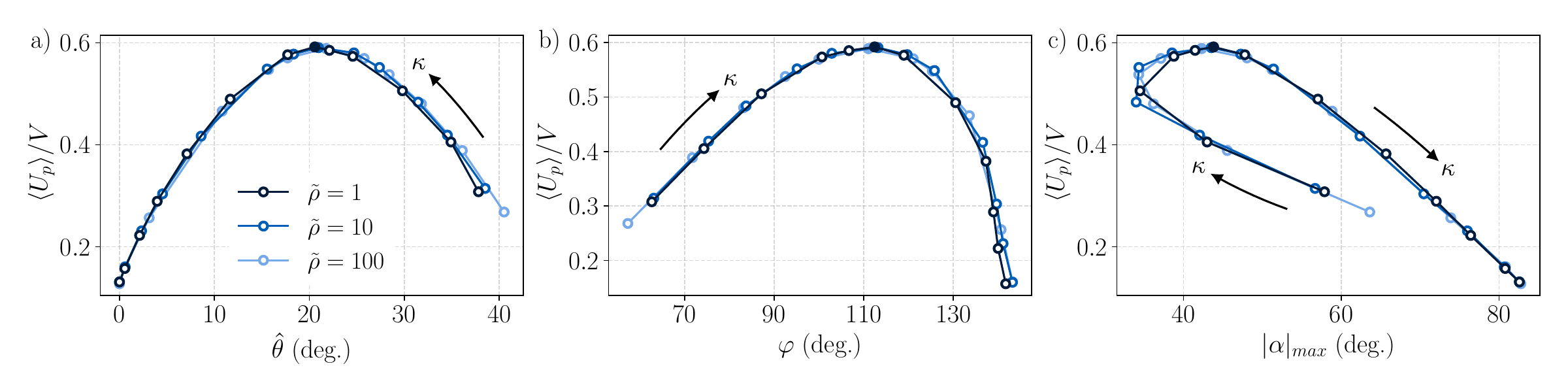}
    \caption{%
    Cycle-averaged propulsive speed $\gls{Up}/\gls{V}$ as a function of
    a) the amplitude \gls{thA} of the first Fourier mode of \gls{th}, 
    b) the phase-lag $\gls{varphi}$ between the heaving motion and the 
    first Fourier mode of $\gls{th}$ and 
    c) the maximum effective angle of attack $\gls{aeA}$.
    The arrows indicate increasing \gls{kn}.
    Filled symbols mark $\gls{kn1}$, $\gls{kn10}$ and $\gls{kn100}$, and
    the legend in a) applies to all panels.}
    \label{fig:res:opt:loops}
\end{figure}

% -----------------------------------------------------------------------------
To compare all simulations, we show in Figure~\ref{fig:res:opt:loops} the
propulsive speed of all stiffnesses and density ratios as a function of
a) the pitching amplitude \gls{thA} of the first Fourier mode of \gls{th},
b) the phase shift \gls{varphi} and c) the maximum effective angle of attack
\gls{aeA}.
In terms of \gls{thA}, the results of the three density ratios collapse onto
a single curve (Figure~\ref{fig:res:opt:loops}a).
Starting from the rigid flapper ($\gls{thA}=0$), the propulsive speed increases 
with the pitching amplitude, reaches its maximum at 
$\gls{thA}=21.1^\circ$ for all three density ratios, and decreases again
for larger amplitudes.
For passively pitching foils of different inertia placed in a uniform stream, 
\citet{phoemsapthawee2020} likewise found the maximum efficiency at the same 
pitching amplitude.
In terms of the phase shift \gls{varphi}, all cases collapse as well
(Figure~\ref{fig:res:opt:loops}b), and the propulsive speed reaches its
maximum at $\gls{varphi}=113^\circ$.
Towards the stiff limit, \gls{varphi} approaches approximately $140^\circ$
and the propulsive speed drops steeply with \gls{varphi}, whereas for softer springs
\gls{varphi} decreases and the propulsive speed decreases only gradually.
For the maximum effective angle of attack in Figure~\ref{fig:res:opt:loops}c), 
the cases of the three density ratios again follow the same path.
\gls{aeA} decreases from $82.5^\circ$ for the rigid flapper to $44^\circ$ at
the optimum, while the propulsive speed increases.
In this range, the flapper feathers more strongly and meets a less inclined
incoming flow, which both reduce \gls{aeA}.
Below the optimum, however, \gls{aeA} first decreases further to 
$34^\circ$ and then increases again for the softest springs, while the
propulsive speed keeps decreasing.
Close to the optimum, the growing pitching amplitude still reduces \gls{aeA}.
For even softer springs, the propulsive speed decreases, so the incoming
flow is more strongly inclined.
In addition, the pitch angle crosses $\gls{th}=0$ only after the stroke
reversal, as shown for the soft case at $\gls{rhor}=1$ ($\gls{kn}=0.047$) in
Figure~\ref{fig:res:single:quat}a).
At the beginning of each half stroke, the flapper is thus still tilted as in
the previous half stroke, and \gls{aeA} increases again.
In contrast to \gls{thA} and \gls{varphi}, the propulsive speed is therefore
not a single-valued function of \gls{aeA}: two flappers with the same \gls{aeA} can
swim at very different speeds, depending on whether their spring is close to
the optimum or much softer.
Flappers with the same pitching kinematics therefore reach the same
propulsive speed, independent of the density ratio.
% -----------------------------------------------------------------------------

\begin{figure}[t]
    \centering
    \includegraphics[width=\textwidth]{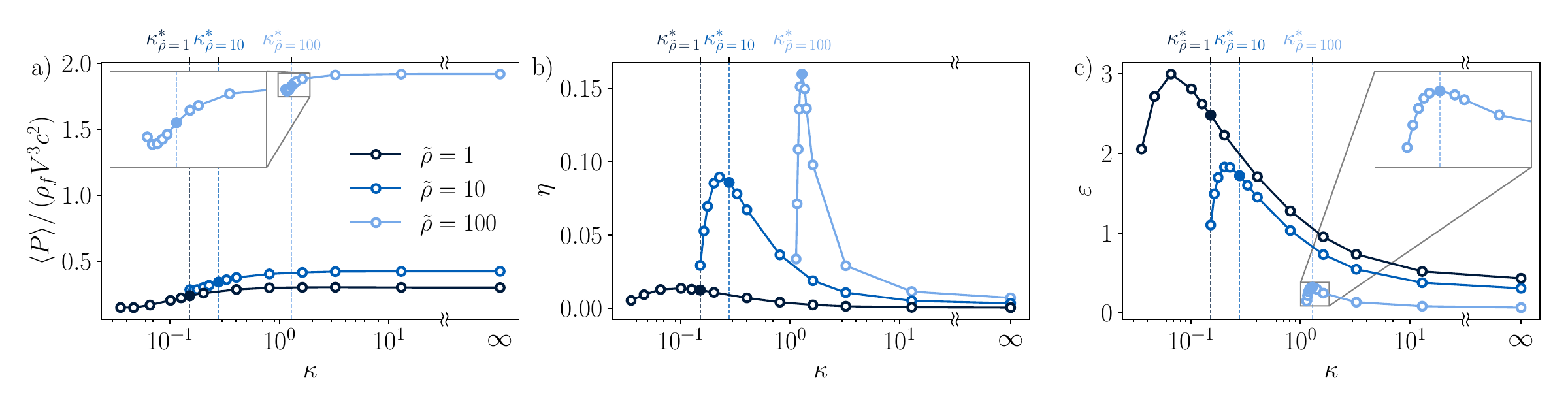}
    \caption{%
    a) Cycle-averaged input power $\gls{P}/(\gls{rhof} \gls{V;cb} \gls{c;sq})$,
    b) propulsive efficiency $\gls{E}$ and c) cruising performance
    $\gls{ve}$ as functions of $\gls{kn}$.
    The insets in a) and c) magnify the region around $\gls{kn100}$.
    The legend in a) applies to all panels, and filled symbols and the
    vertical dashed lines mark $\gls{kn1}$, $\gls{kn10}$ and $\gls{kn100}$.
    The break in the abscissa indicates that the rigid case is not drawn
    to scale.
    }
\label{fig:res:flapper:opt:power}
\end{figure}

% -----------------------------------------------------------------------------
%   III - Power and Efficiencies
% -----------------------------------------------------------------------------
Finally, Figure~\ref{fig:res:flapper:opt:power} reports the input power and the 
resulting performance.
For all three density ratios, \gls{P} follows the trend found for 
$\gls{rhor}=1$: it increases with \gls{kn} and levels off towards the rigid
flapper (Figure~\ref{fig:res:flapper:opt:power}a).
The level of \gls{P}, however, increases strongly with the density ratio,
from about $0.3$ for $\gls{rhor}=1$ to about $1.9$ for $\gls{rhor}=100$ in
the rigid limit.
This difference comes from the inertia of the flapper: a flapper of larger mass
requires more power to be accelerated during the heaving motion.
The rigid flapper nevertheless reaches the same propulsive speed for all three 
density ratios (Figure~\ref{fig:res:opt:Up}).
For $\gls{rhor}=100$, the inertia of the flapper outweighs the hydrodynamic
forces, so \gls{P} varies by less than $10\%$ over the whole range of
\gls{kn}.
For $\gls{rhor}=1$, in contrast, \gls{P} doubles from the softest spring to
the rigid flapper.
The inset of Figure~\ref{fig:res:flapper:opt:power}a) magnifies \gls{P}
around \gls{kn100}.
For $\gls{rhor}=10$ and $100$, \gls{P} reaches its minimum below the optimum
and rises again for the softest springs.
This rise is associated with an additional peak of \gls{ae} slightly after
stroke reversal found for the softest springs, as discussed for the case with
$\gls{rhor}=1$ in \S~\ref{sec:results:kinematics}.
% -----------------------------------------------------------------------------

% -----------------------------------------------------------------------------
Since the three flappers reach the same propulsive speed at the optimum 
\gls{kn}, the density ratio affects the performance indicators only through 
the input power.
The cruising performance \gls{ve}, which does not contain the flapper mass,
therefore decreases strongly with the density ratio
(Figure~\ref{fig:res:flapper:opt:power}c).
A flapper of higher density ratio thus travels a shorter distance per unit
input work, as also found by \citet{arora2018} for a plunging plate with a
torsional spring.
The propulsive efficiency \gls{E} shows the opposite ordering
(Figure~\ref{fig:res:flapper:opt:power}b), but only because it relates the
kinetic energy of the flapper to the input power \citep{arranz2022}.
From $\gls{rhor}=1$ to $100$, the mass of the flapper increases by a factor
of $100$, whereas \gls{P} at the optimum increases only by a factor of about
$7.5$.
\gls{E} is therefore not suited to compare flappers of different density
ratios, and the comparison relies on \gls{ve}.
The stiffness at which \gls{E} and \gls{ve} are maximal lies below
\gls{knOPT} for $\gls{rhor}=1$ and approaches it as the density ratio
increases.
For $\gls{rhor}=1$, \gls{ve} is maximal well below \gls{kn1}, because for
softer springs the input power decreases faster than the propulsive speed.
For $\gls{rhor}=100$, \gls{P} varies by less than $10\%$ with \gls{kn}, and
both \gls{ve} and \gls{E} are maximal at \gls{kn100}
(Figure~\ref{fig:res:flapper:opt:power}b and the inset of
Figure~\ref{fig:res:flapper:opt:power}c).
The higher the density ratio, the less the choice of the spring involves a
trade-off between propulsive speed and energetic performance.
For all three density ratios, the maxima of \gls{E} and \gls{ve} also lie
at springs stiffer than $\kappa_{res}$.
% -----------------------------------------------------------------------------

%% file: tex/04_conclusion.tex
% =============================================================================
% =============================================================================
%   CONCLUSION
% =============================================================================
% =============================================================================
\section{Conclusions} \label{sec:conclusions}
In this work, we have investigated numerically a three-dimensional
flapper that self-propels from a prescribed sinusoidal heaving motion of its 
centre of mass.
It is modelled as a rigid body that pitches passively about a spanwise axis,
restrained by a torsional spring.
We have varied the normalised spring stiffness \gls{kn} for the three density
ratios $\gls{rhor}=1$, $10$ and $100$ to evaluate its influence on the propulsive
performance of the flappers.
For each density ratio we have further optimised the value of \gls{kn} that
maximises propulsive speed.
% -----------------------------------------------------------------------------

% -----------------------------------------------------------------------------
For the cases with $\gls{rhor}=1$, the maximum propulsive speed achieved is
$\gls{Up}/\gls{V} = 0.59$ and the normalised stiffness $\gls{kn}=0.15$.
The passive pitching motion obtained when the stiffness of the spring is 
the optimum results in an increase of the propulsive speed of roughly $350\%$
as compared with the non-pitching case.
This enhancement of propulsive speed is accompanied by a reduction of the input
power of approximately $20\%$, making the passive pitching implemented in this
work an attractive strategy to improve propulsive performance.
Please note that this improvement is notably lower than the one obtained with a 
chordwise-flexible plate presented in \citet{arranz2022} with otherwise the same
computational setup.
Nevertheless, the solution proposed here is simpler and because the location 
of the spring is at the centre of mass instead of the leading edge, the centre 
of mass does not actuate as a lumped mass that would result in larger structural 
demands. 
This is, however, only a conjecture and should be explored in detail.
% -----------------------------------------------------------------------------

% -----------------------------------------------------------------------------
If we consider a hypothetical reference feathering case in which stroke 
reversal occurs when the flapper is aligned with the free stream ($\gls{th}=0$),
we find that the optimum case results from a balance between leading,
small-amplitude, pitching motions associated with stiffer springs, and 
lagging, large-amplitude, pitching motions associated with softer springs.
Feathering substantially improves the propulsive performance of the flapper 
when it acts as a damper of the large effective angles of attack (that would 
otherwise occur in pure heaving), in addition to transferring part of the normal
force (typically originated with circulatory effects) to the forward direction.
Stiffer springs show small pitching angles, therefore the benefits from
feathering are not felt by the flapper: the effective angles of attack are
still too large, flow separation occurs at both the leading and trailing edges
and the hydrodynamic force acts mainly in the vertical direction.
Softer springs, however, show large pitching angles, but the phase difference
between the pitching motion and the heaving motion is smaller than $90^\circ$,
which means that the pitching motion now lags with respect to the reference
feathering motion and the flapper is horizontally aligned ($\gls{th}=0$) after
stroke reversal.
This late turn of the flapper results in an additional peak of the effective angle 
of attack slightly after stroke reversal, which turns out to affect negatively 
the propulsion performance of the flapper.
The optimum is then found as a balance between these two limits, both in the
pitching amplitude and in the phase difference between the heaving and pitching
motions.
More specifically, the optimum case results in a pitching amplitude (first 
Fourier mode coefficient) of $21.1^\circ$, and the pitching motion leads the 
heaving motion by $113^\circ$, or equivalently, leads by $23^\circ$ the 
reference feathering motion (the flapper is horizontally aligned before stroke
reversal).
% -----------------------------------------------------------------------------

% -----------------------------------------------------------------------------
Interestingly, all three density ratios reach the same maximum propulsive speed,
but the higher the density ratio, the stiffer the spring needs to be to reach it.
There is a factor of $8.5$ between the normalised stiffness obtained for the
case with $\gls{rhor}=100$ and that obtained for the case with $\gls{rhor}=1$. 
When the propulsive speed is represented versus the normalised stiffness shifted
by the respective optimum, the results for the three density ratios collapse onto
a single curve.
This collapsing of the results is observed not only in cycle-averaged quantities
but also in the instantaneous evolution of the pitching motion, effective angle of 
attack and vertical hydrodynamic force.
Furthermore, we found that for all three density ratios, the spring that maximises
the propulsive speed is stiffer than the spring at which the flapper would 
resonate in quiescent fluid, including the added inertia of the fluid in the 
analysis.
The difference between the optimum and the resonant stiffness is nearly
independent of the density ratio.
% -----------------------------------------------------------------------------

% -----------------------------------------------------------------------------
The input power, however, increases strongly with the density ratio because
of the inertia of the flapper.
The propulsive efficiency \gls{E} shows that a flapper of higher density ratio
travels a shorter distance per unit input work, whereas the mass-specific 
quantity \gls{ve} (cruising performance) decreases with the density ratio.
The stiffness at which \gls{ve} and \gls{E} are maximal lies below
\gls{knOPT} for $\gls{rhor}=1$ and approaches it as the density ratio
increases.
The higher the density ratio, the less the choice of the spring involves a
trade-off between propulsive speed and energetic performance.
% -----------------------------------------------------------------------------

% -----------------------------------------------------------------------------
Overall, the optimal spring is not set by resonance but by the feathering
motion it produces.
For all density ratios considered, the highest propulsive speed is reached
when the flapper feathers with an amplitude of $21.1^\circ$ and leads the
reference feathering motion, and a flapper of higher density ratio only needs a
stiffer spring to reach this motion.
% -----------------------------------------------------------------------------

%% file: tex/05_appendix.tex
% =============================================================================
% =============================================================================
%   APPENDIX
% =============================================================================
% =============================================================================
\section{Appendix} \label{sec:app}

% -----------------------------------------------------------------------------
%   Flapper Geometry
% -----------------------------------------------------------------------------
\subsection{Flapper geometry}
\label{sec:app:flapper_geometry}

% -----------------------------------------------------------------------------
\begin{figure}[h!]
    \centering
    \includegraphics[width=0.8\textwidth]{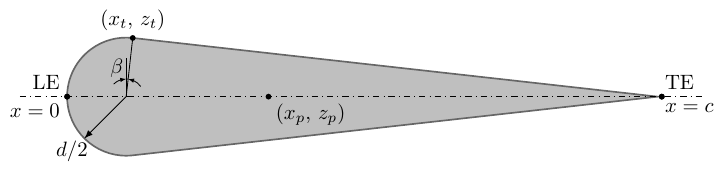}
    \caption{Cross-section of the three-dimensional flapper: a circular leading
        edge of diameter $\gls{d}$ blended into two straight edges meeting at 
        the trailing edge $x=\gls{c}$, with tangent point $(x_t,z_t)$, edge 
        angle $\beta$, and centre of mass $(\gls{vxp_x},\gls{vxp_z})$.}
    \label{fig:app:flapper}
\end{figure}
The flapper cross-section is shown in Figure~\ref{fig:app:flapper}: a circular
leading edge of diameter $\gls{d}$ that blends tangentially into two straight 
edges converging at the trailing edge. With $x$ measured from the leading
edge, the half-thickness profile is
\begin{equation}
    f(x) =
    \begin{cases}
        \sqrt{(\gls{d}/2)^2 - (x - \gls{d}/2)^2}, & 0 \le x \le x_t \,, \\[4pt]
        (\gls{c} - x)\tan\beta, & x_t \le x \le \gls{c} \,,
    \end{cases}
    \label{eq:app:profile}
\end{equation}
where the first branch describes the circular leading edge and the second the
straight edge. The two branches join at the tangent point
$x_t = \tfrac{\gls{d}}{2}(1 + \sin\beta)$, and requiring the straight edge to be
tangent to the circle fixes the edge angle 
$\beta = \arcsin[\gls{d}/(2\gls{c} - \gls{d})]$.
% -----------------------------------------------------------------------------

% -----------------------------------------------------------------------------
The area and chordwise centroid are
\begin{align}
    A_c &= 2 \int_0^{\gls{c}} f \, \mathrm{d}x
       = 2 \int_0^{x_t} \sqrt{(\gls{d}/2)^2 - (x - \gls{d}/2)^2} \, \mathrm{d}x
       + 2 \int_{x_t}^{\gls{c}} (\gls{c} - x)\tan\beta \, \mathrm{d}x \nonumber
        \\[4pt]
      &= \frac{\gls{d}^2}{4}
         \left( \beta + \frac{\pi}{2} + \frac{\cos\beta}{\sin\beta} \right)
      % = \frac{d^2}{4}
      %   \left( \beta + \frac{\pi}{2} + \cot\beta \right) ,
    \label{eq:app:area} \\[6pt]
    \gls{vxp_x} &= \frac{2}{A_c} \int_0^{\gls{c}} x\,f \, \mathrm{d}x
         = \frac{2}{A_c} \left[
            \int_0^{x_t} x\,\sqrt{(\gls{d}/2)^2 - (x - \gls{d}/2)^2} \, \mathrm{d}x
          + \int_{x_t}^{\gls{c}} x\,(\gls{c} - x)\tan\beta \, \mathrm{d}x \right] \nonumber
        \\[4pt]
        &= \frac{\gls{d}}{\beta+\frac{\pi}{2}+\frac{\cos\beta}{\sin\beta}}
           \left[\frac{\beta}{2}+\frac{\pi}{4}+\frac{\sin\beta\cos\beta}{2}
             -\frac{\cos^3\beta}{3}
             +\frac{\cos^3\beta\left(3+3\sin\beta
                 -2\cos^2\beta\right)}{6\sin^2\beta}\right]
    \label{eq:app:centroid}
\end{align}
with $\gls{vxp_z} = 0$ by symmetry, and the volume follows by extruding the
cross-section over the span, $\gls{Vp} = A_c\,\gls{b}$.
% -----------------------------------------------------------------------------

% -----------------------------------------------------------------------------
All moments below are taken about the spanwise \gls{vx_y}-axis, the only axis
about which the flapper rotates under the planar-motion constraint of
\S\ref{sec:meth:prob}.
Integrating $(x^2 + z^2)$ over the cross-section and extruding over the
span, then splitting at $x_t$, gives the moment of inertia about the
leading edge,
\begin{align}
    I_{y,\mathrm{(LE)}}
    &= \gls{rhop} \gls{b} \int_{A_c} \!\left(x^2 + z^2\right) \mathrm{d}A
    = 2 \gls{rhop} \gls{b} \int_0^{\gls{c}} \!\left( x^2 f + \tfrac{1}{3} f^3 \right)
      \mathrm{d}x \nonumber
      \\[4pt]
    &= 2 \gls{rhop} \gls{b} \int_0^{x_t} \!\left(
        x^2 \sqrt{(\gls{d}/2)^2 - (x - \gls{d}/2)^2}
        + \tfrac{1}{3} \left[(\gls{d}/2)^2 - (x - \gls{d}/2)^2\right]^{3/2}
      \right) \mathrm{d}x \nonumber
      \\[4pt]
    &\quad + 2 \gls{rhop} \gls{b} \int_{x_t}^{\gls{c}} \!\left(
        x^2 (\gls{c} - x)\tan\beta
        + \tfrac{1}{3} (\gls{c} - x)^3 \tan^3\beta
      \right) \mathrm{d}x \nonumber
      \\[4pt]
    &= \frac{\gls{rhop} \gls{b}\,\gls{d}^4}{192}
      \left[\, 9(\pi + 2\beta) + \frac{18\cos\beta}{\sin\beta}
      +\frac{8\cos^3\beta}{\sin^2\beta}+\frac{2\cos^3\beta}{\sin^3\beta} \,\right]
    \label{eq:app:ILEc}
\end{align}
and the corresponding value at the centre of mass follows from the
parallel-axis theorem,
$I_{y,(\gls{vxp_x})} = I_{y,\mathrm{(LE)}} - \gls{rhop} \gls{Vp}\,\gls{vxp_x}^2$.
The specific moment of inertia, i.e.\ the moment of inertia per unit mass,
is then
\begin{align}
   \gls{J} &= \frac{I_{y,(\gls{vxp_x})}}{\gls{rhop} \gls{Vp}}
        = \frac{\gls{d}^2 \left[\, 9(\pi + 2\beta)
          + \dfrac{18\cos\beta}{\sin\beta}
          + \frac{8\cos^3\beta}{\sin^2\beta}+\frac{2\cos^3\beta}{\sin^3\beta} \,\right]}
          {48 \left( \beta + \dfrac{\pi}{2}
          + \dfrac{\cos\beta}{\sin\beta} \right)}
          - \gls{vxp_x}^2
    \label{eq:app:Jy}
\end{align}
In this work we set the diameter to $\gls{d}=0.2\gls{c}$ and the span to
$\gls{b}=0.5\gls{c}$, so that equations~\eqref{eq:app:area}--\eqref{eq:app:Jy} give
$A_c = 0.106\,\gls{c;sq}$, $\gls{vxp_x} = 0.349\,\gls{c}$ and 
$\gls{J} = 0.055\,\gls{c;sq}$, and a volume of the flapper 
$\gls{Vp} = A_c\,\gls{b} = 0.053\,\gls{c;cb}$.
% -----------------------------------------------------------------------------

% -----------------------------------------------------------------------------
%   Added moment of inertia
% -----------------------------------------------------------------------------
\subsection{Added moment of inertia}
\label{sec:app:added_inertia}
We estimate the added moment of inertia $I_a$ following \citet{brennen1982},
for a thin plate of chord \gls{c} and span \gls{b} in potential flow.
For a plate that is symmetric fore and aft, a rotation with angular velocity
\gls{om} about \gls{vxp_x} is equivalent to a rotation about the mid-chord
and a heaving motion of the mid-chord with velocity
$(\gls{c}/2-\gls{vxp_x})\gls{om}$, so that
\begin{equation}
    I_a = I_{a,m} + \left(\frac{\gls{c}}{2}-\gls{vxp_x}\right)^2 m_a,
    \label{eq:app:Ia_split}
\end{equation}
where $I_{a,m}$ is the added moment of inertia about the mid-chord and $m_a$
the added mass in heave.
With strip theory, i.e.\ the values of a two-dimensional plate
\citep[Table~II]{brennen1982} multiplied by the span, these are
\begin{equation}
    I_{a,m} = \frac{\pi}{8}\,\gls{rhof}\left(\frac{\gls{c}}{2}\right)^4\gls{b},
    \qquad
    m_a = \pi\,\gls{rhof}\left(\frac{\gls{c}}{2}\right)^2\gls{b}.
    \label{eq:app:strip}
\end{equation}
Since strip theory neglects the flow around the tips, we correct both
contributions with an elliptic disk of the same chord and span, for which
exact three-dimensional values are available \citep[Table~IV]{brennen1982}.
The ratio between the exact and the strip-theory value of the disk is
$0.655$ for the rotation and $0.413$ for the heave, and we assume the same
ratios for the flapper.
For the heave, measurements on rectangular plates
\citep[Table~III]{brennen1982} give $0.420$, within $2\%$ of the disk value.
With $\gls{b}=0.5\gls{c}$ and $\gls{vxp_x}=0.349\gls{c}$, this gives
$I_a\approx0.0117\,\gls{rhof}\gls{c}^5$.
% -----------------------------------------------------------------------------

% -----------------------------------------------------------------------------
%   Grid sensitivity analysis
% -----------------------------------------------------------------------------
\subsection{Grid sensitivity analysis} \label{sec:app:gridstudy}

\begin{figure}[b!]
    \centering
    \includegraphics[width=1.0\textwidth]{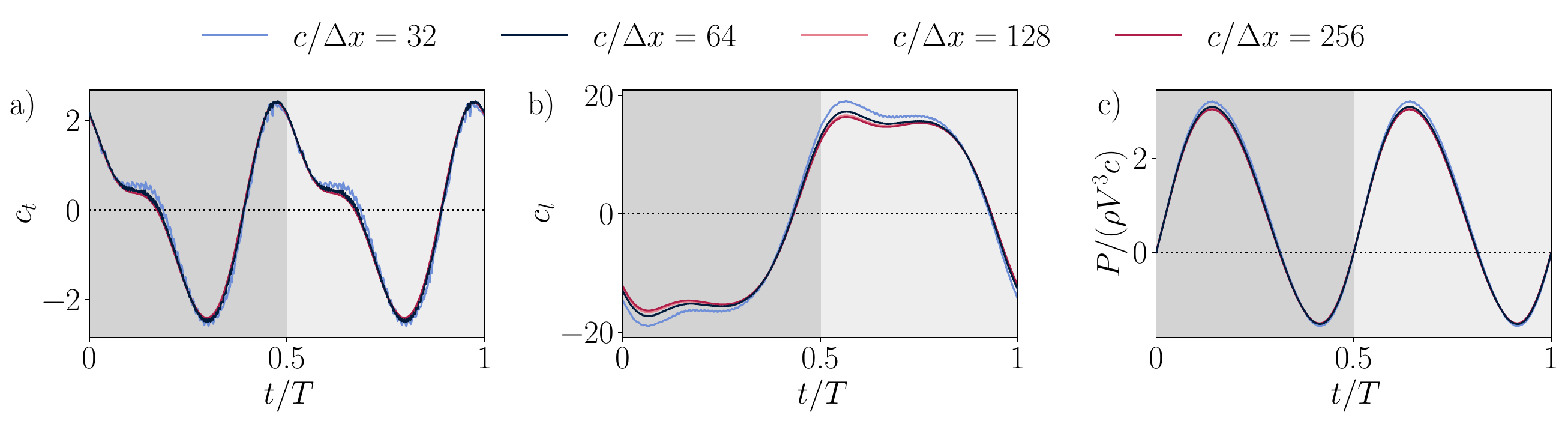}
    \caption{Grid sensitivity analysis on the a) thrust
    coefficient $c_t$, b) lift coefficient $c_l$ and c) input power per unit 
    span $\gls{P_i}$ over one flapping cycle of an isolated flapper with a
    prescribed motion of its centre of mass.
    The dark and light shaded regions mark the downstroke
    ($0 \le t/\gls{T} \le 0.5$) and the upstroke ($0.5 \le t/\gls{T} \le 1$),
    respectively.}
    \label{app:force}
\end{figure}

\begin{figure}[t!]
    \centering
    \includegraphics[width=0.5\textwidth]{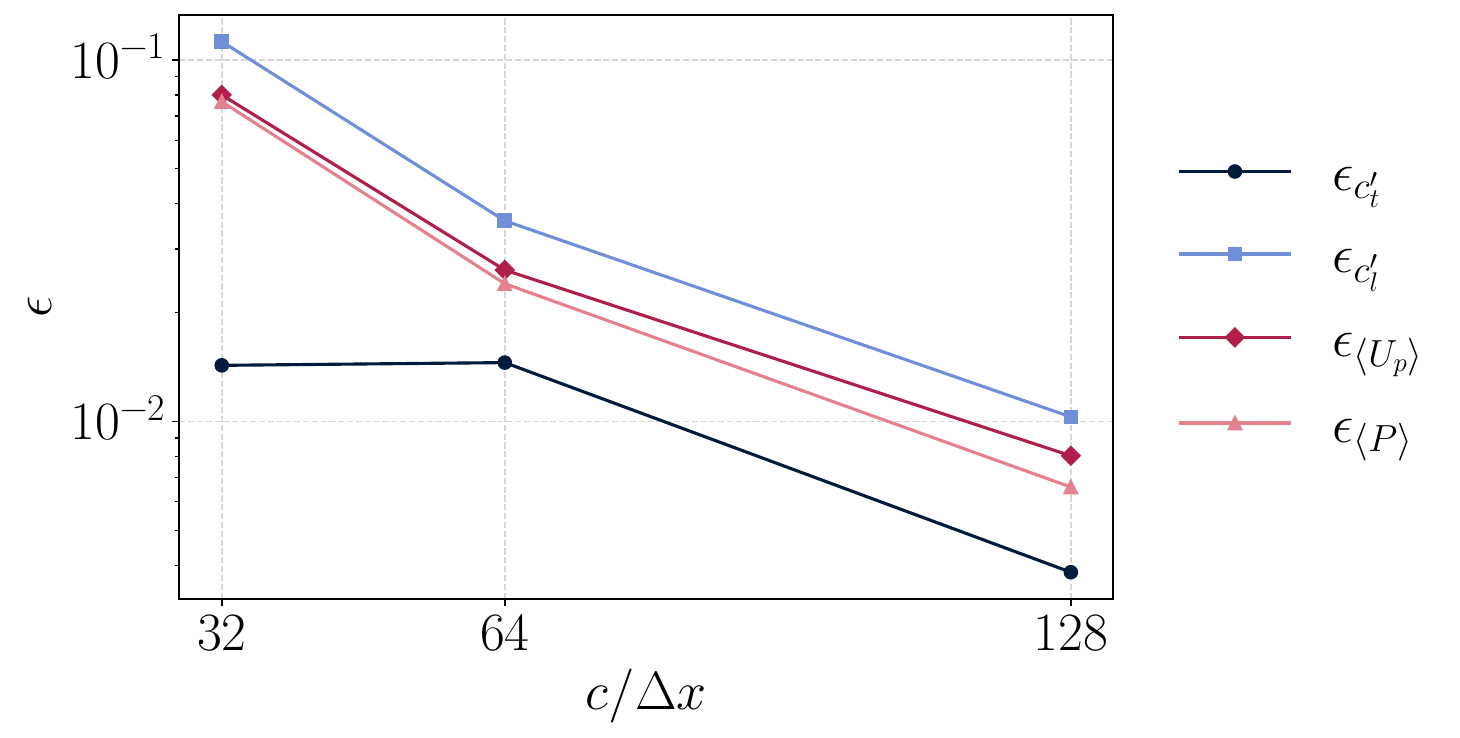}
    \caption{Relative errors of the four quantities used to assess
    convergence --- the root-mean-square of the thrust and lift
    coefficients, the cycle-averaged input power and the propulsive speed
    --- with respect to the finest resolution ($\gls{c}/\Delta x = 256$).}
    \label{app:error}
\end{figure}

% -----------------------------------------------------------------------------
In this appendix we present the grid refinement study carried out to select
the spatial resolution used in the simulations.
While the simulations of the main paper are three-dimensional, the
refinement study is performed on a simplified two-dimensional domain of the
same streamwise and vertical extent ($L_x=16\gls{c}, \, L_z=8\gls{c}$).
Since the flapper motion in the main paper is already planar --- streamwise
translation through the fluid--structure coupling, imposed heaving, and
pitching about the spanwise axis --- the reduction to two dimensions
discards no relevant dynamics while considerably lowering the computational
cost of the higher-resolution cases.
The grid refinement study is performed for a 2D set-up of the flapper
configuration.
It is carried out at the same governing parameters provided
in~\S~\ref{sec:meth:prob}, namely $\gls{ReV}=200$ and heaving
amplitude $\gls{A}=0.5\gls{c}$, while the same heaving motion (see 
equation~\eqref{eq:meth:pos}) is imposed.
Since the grid refinement study is not concerned with the optimisation of
the torsional spring stiffness, the pitching motion is, for simplicity, imposed
directly as
\begin{equation}
\gls{th}(t) = \gls{th}_0 \cos(2\pi \gls{f} t + \gls{varphi}),
\end{equation}
with pitching amplitude $\gls{th}_0 = \pi/6$ and a phase shift
$\gls{varphi} = \pi/2$ between the pitching and heaving motions, the latter
chosen following~\citet{moriche2017}, who reported maximum propulsive
efficiency at this phase shift.
Four successively refined grids are considered, with $\gls{c}/\Delta x = 32$,
$64$, $128$ and $256$ points per chord and a constant refinement ratio of
two.
The temporal resolution is refined together with the grid, using
$S_{pp} = 1024$, $2048$, $4096$ and $8192$ time steps per period for the
four grids, respectively, i.e.\ $\Delta t = \gls{T}/S_{pp}$ ($\gls{T}=1/\gls{f}$).
This keeps $CFL = U_{\mathrm{max}} \Delta t/\Delta x \lesssim 0.3$ on all
grids, with $U_{\mathrm{max}}$ the maximum flow velocity in the domain.
Because the flapper is self-propelled and the simulation is carried out in a
moving frame, the inflow velocity $U$ is iterated, as described
in~\S~\ref{sec:meth:comp} and in detail in~\citet{moriche2026}, until it
matches the propulsive velocity up to convergence, so that the flapper is 
stationary on cycle-average  in the domain.
At steady state the cycle-averaged streamwise force therefore vanishes, and
the cycle-averaged vertical force vanishes by symmetry of the imposed motion
about the streamwise axis, as shown in Figure~\ref{app:force}b).
Hence, the convergence is assessed on the four quantities:
the root-mean-square of the thrust and lift coefficients, the
cycle-averaged  propulsive speed, $\gls{Up}$ (see
equation~\eqref{eq:meth:velocity}) and cycle-averaged input power, $\gls{P}$.
The thrust and lift coefficients are defined as
\begin{equation}
c_t = \frac{-2\gls{vF_x}}{\gls{rhof} \gls{c;cb} \gls{f}^2},
\qquad
c_l = \frac{2\gls{vF_z}}{\gls{rhof} \gls{c;cb} \gls{f}^2},
\end{equation}
where $\gls{vF_x}$ and $\gls{vF_z}$ denote the streamwise and vertical 
components of the fluid force acting on the flapper.
Since the flow is fully periodic, Figure~\ref{app:force} additionally shows
the instantaneous thrust coefficient, lift coefficient and input power over
the last flapping cycle.
Note that the instantaneous input power in Figure~\ref{app:force}c) takes
negative values during part of the cycle, whereas these are excluded from the
cycle-averaged value by the clipping to positive power 
(see equation~\eqref{eq:meth:IP}).
To quantify the convergence, we define for a generic quantity $\phi$ the
relative error with respect to the finest resolution,
\begin{equation}
\epsilon_\phi(r) =
\frac{\left| \phi_r - \phi_{256} \right|}{\left| \phi_{256} \right|},
\end{equation}
where $r$ denotes the resolution $\gls{c}/\Delta x$ of each case, and $\phi$ stands
in turn for $c'_t$, $c'_l$ (prime indicating the root-mean-square value), the
cycle-averaged propulsive velocity and cycle-averaged input power.
While some deviations are observed for $\gls{c}/\Delta x = 32$, all four quantities
converge towards the highest-resolution case as the resolution is increased.
As shown in Figure~\ref{app:error}, the errors in all four quantities are
already reasonably low at $\gls{c}/\Delta x = 64$.
This resolution is therefore adopted in our calculations, as a compromise
between computational cost and accuracy.
% -----------------------------------------------------------------------------